\documentclass[12pt]{article} 

\usepackage[usenames]{xcolor}
\usepackage{setspace}

\usepackage{graphicx}
\usepackage{amsmath}
\usepackage{amssymb}
\usepackage{amsthm}

\usepackage{hhline}
\usepackage{array}
\newcolumntype{I}{!{\vrule width 3\arrayrulewidth}}
\newcommand\whline{%
    \noalign{\xdef\origarrayrulewidth{\the\arrayrulewidth}%
    \global\arrayrulewidth 3\arrayrulewidth}%
    \hline%
    \noalign{\global\arrayrulewidth\origarrayrulewidth}%
}
\newcolumntype{V}{!{\vrule width \arrayrulewidth}}
\newlength\savedwidth 

\makeatletter
\newcommand{\ArrayVdashLine}{%
  \raisebox{-\dp\@arstrutbox}{%
    \vbox to \dimexpr\ht\@arstrutbox+\dp\@arstrutbox\relax{%
      \offinterlineskip
      \cleaders\vbox{%
        \hrule width \arrayrulewidth height 4pt
        \kern 3pt
      }\vfill
    }%
  }%
}
\makeatother
\newcolumntype{D}{!{\ArrayVdashLine}}
\usepackage{bm}
\usepackage {arydshln}
\usepackage{ulem}%\uline{} math OK., \uuline,uwave,sout,xout,

\newcommand{\MyDecom}[6]{(\repr{#1},\repr{#2})_{#3,#4,#5}^{#6}}
\newcommand{\Eqref}[1]{eq.~(\ref{#1})}
\newcommand{\MyDecomC}[8]{(\repr{#1},\repr{#2},\repr{#3})_{#4,#5,#6,#7}^{#8}} 
 
\newcommand{\VEV}[1]{\left\langle #1 \right\rangle}

\newcommand{\nn}{\nonumber}

\newcommand{\repr}[1]{{\bf#1}} 
\newcommand{\sub}[1]{_{\rm{#1}}}

\newcommand{\Z}[1]{{\mathbb Z}_#1}

\newcommand{\bequ}{\begin{equation}}
\newcommand{\eequ}{\end{equation}}
\newcommand{\beqn}{\begin{eqnarray}}
\newcommand{\eeqn}{\end{eqnarray}}
\newcommand{\bctr}{\begin{center}}
\newcommand{\ectr}{\end{center}}
\newcommand{\bit}{\begin{itemize}}
\newcommand{\eit}{\end{itemize}}
\newcommand{\Ls}{\left(}
\newcommand{\Rs}{\right)}

\allowdisplaybreaks[4]

\usepackage{multirow}
\numberwithin{equation}{section}
\begin{document}

\begin{flushright}
\today
\end{flushright}

\begin{center}

{\LARGE\bf SU(9) grand unified model with rank-reducing discrete boundary conditions on $T^2/\Z4$}

\vskip 1.4cm

{\large  
$^{a}$Yoshiharu Kawamura\footnote{e-mail:haru@azusa.shinshu-u.ac.jp},
$^{b}$Kentaro Kojima\footnote{e-mail:kojima@artsci.kyushu-u.ac.jp}\\
and\\
$^{c}$Toshifumi Yamashita\footnote{e-mail:tyamashi@aichi-med-u.ac.jp}
}
\\
\vskip 1.0cm
{\it $^a$Department of Physics, Shinshu University, Matsumoto 390-8621, Japan}\\
{\it $^b$ Faculty of Arts and Science, Kyushu University, Fukuoka 819-0395, Japan}\\
{\it $^c$Department of Physics, Aichi Medical University, Nagakute 480-1195, Japan}
\vskip 1.0cm

\begin{abstract}
We study a six-dimensional SU(9) grand unified model with rank-reducing discrete boundary conditions and continuous Wilson line phases on the orbifold $T^2/\Z4$.
    We show that the model can realize grand unification of the Standard Model gauge interactions and electroweak symmetry breaking via the Hosotani mechanism.
    The model has several attractive features. Two Higgs doublets arise as zero modes of the extra-dimensional components of the gauge field, and the leptons and quarks in each generation are separately embedded into two bulk multiplets, in the $\bm{36}$ and $\bm{84}$ representations of SU(9), respectively, without exotic matter.
   Furthermore, proton decay processes mediated by the gauge bosons in the bulk are absent, because the leptons and quarks belong to different bulk multiplets.
\end{abstract}

\end{center}

\vskip 1.0 cm

%%%%%%%%%%%%%%%%%%%%%%%%%%%%%%%%%%%%%%%%%%%%%%%%%%%%%%%%%%%%%%%%
%
\section{Introduction}
%
%%%%%%%%%%%%%%%%%%%%%%%%%%%%%%%%%%%%%%%%%%%%%%%%%%%%%%%%%%%%%%%%

Gauge theories in higher-dimensional space-time provide an attractive
framework for physics beyond the Standard Model (SM).  For instance,
orbifold compactification can realize chiral fermions and nontrivial
gauge symmetry breaking through boundary conditions (BCs).  In
particular, the latter can be an elegant
solution~\cite{K1,K2,H&N,Hebecker:2001jb} to the so-called
doublet-triplet splitting problem in the grand unified theories
(GUTs)~\cite{GUT,SUSYGUT-DG,SUSYGUT-S}.

Another important idea is gauge-Higgs unification, in which
four-dimensional (4D) Higgs fields originate from the
extra-dimensional components of higher-dimensional gauge
fields~\cite{M}.  In such models, the Higgs fields may be related to
Wilson line phases, and electroweak symmetry breaking (EWSB) can be
induced dynamically by the Hosotani
mechanism~\cite{H1,H2,GHU-KL&Y,HHH&K,HH&K}.  Although the Higgs sector
may also contain non-flat degrees of freedom (d.o.f.) that are not
identified with Wilson line phases, the Wilson line d.o.f.  themselves
are non-local and therefore the Wilson-line-dependent part of the
effective potential is finite without supersymmetry
(SUSY)~\cite{finiteness-K,finiteness-HI&L,finiteness-AC&G}.  The local
divergences in subdiagrams of the Wilson-line effective
potential~\cite{finiteness-M&Y,finiteness-HMT&Y,divcont} are removed
by the lower-loop counter terms, without introducing independent
counter terms~\cite{finiteness-M&Y,finiteness-HMT&Y}.  This idea
works, in addition to EWSB~\cite{M,GHU-KL&Y,GHU-CG&M,GHU-SS&S}, also
in the GUT
context~\cite{GHU-HHK&Y,Lim:2007jv,Hosotani:2015hoa,Maru:2019lit}.
Furthermore, the unification of the GUT-breaking Higgs field with the
unified gauge field has been
studied~\cite{gGHU-E6,gGHU,gGHU-Y,gGHU-KT&Y,
  Kojima:2023mew,gGHU-pheno,gGHU-pheno-NSS&Y}.

Orbifold BCs are specified by twist matrices acting on the internal
space of gauge and matter fields.  In many familiar orbifold models,
these twist matrices can be chosen simultaneously diagonal, and the
rank of the gauge group is not reduced by the BCs, if
continuous d.o.f. of Wilson line phases does not acquire
nontrivial vacuum expectation values (VEVs)~\cite{S&S,FN&W}.  However,
as shown in ref.~\cite{Kawamura:2022ecd}, this is not always the case
on $T^2/\Z4$ and $T^2/\Z6$.  On these orbifolds, twist matrices can
contain non-diagonal blocks which cannot be simultaneously
diagonalized by unitary and gauge transformations.  Such non-diagonal
discrete BCs can reduce the rank of the gauge symmetry without
nontrivial VEVs of the Wilson line phase.  This provides a new class
of symmetry-breaking mechanisms based on discrete Wilson line phases.

In our previous work \cite{Kawamura:2025}, we studied six-dimensional
(6D) SU($n$) gauge models with rank-reducing discrete BCs on
$T^2/\Z4$.  In particular, we found that an SU(6) gauge model, which
is the minimal model when we require also continuous Wilson line
phases, realizes an interesting electroweak model.  In this model, two
Higgs doublets arise from the zero modes of the extra-dimensional
components of the gauge
field~\cite{Antoniadis:2001cv,Chang:2012iq,Matsumoto:2014ila,Hasegawa:2015vqa,Akamatsu:2023ird,Akamatsu:2026sjg},
and the quarks in each generation can be obtained from a bulk fermion
in the $\bm{15}$ representation of SU(6) without exotic zero modes.
The EWSB can also be realized by the Hosotani mechanism for suitable
bulk matter contents.

The SU(6) model, however, 
requires a 
large representation to serve the leptons as zero modes, 
and predicts an excessively large value of the Weinberg angle at the compactification scale. 
As is well known, the Weinberg angle can be adjusted by introducing additional 
U(1) gauge symmetry, which may be also used for obtaining leptons
from bulk zero modes in a smaller representation. 
Including the color gauge symmetry, then the gauge symmetry would 
be SU(6)$\times$SU(3)$\times$U(1), which motivates SU(9) GUTs. 

In this paper, we construct a 6D SU(9) grand unified model on
$T^2/\Z4$ with rank-reducing discrete BCs and continuous Wilson line
phases.  The quarks and leptons in each generation arise from bulk
fermions in the $\bm{84}$ and $\bm{36}$ representations of SU(9),
respectively, and surprisingly without exotic zero modes.  The two
Higgs doublets originate from the extra-dimensional components of the
gauge field, as in the SU(6) model.  Depending on the 6D chiralities
of the bulk fermions, both type-II and type-Y (flipped) two Higgs
doublet models (2HDMs) ~\cite{BH&P,Grossman,AKT&Y,Branco:2011iw} can
be obtained as low-energy effective theories.

We also discuss several consistency and phenomenological aspects of
the model.  We examine bulk anomaly cancellation, gauge coupling
relations, proton stability and the EWSB.  In particular, proton decay
mediated by bulk gauge bosons is absent because quarks and leptons
belong to different bulk multiplets.  We also study the one-loop
effective potential for the Wilson line phases and show, by adding
suitable bulk matter fields, that EWSB vacua can be obtained.  The
numerical examples found in this paper should be regarded as toy
models: the compactification scale is about $10$--$100$ times larger
than the electroweak scale, while the gauge coupling unification
indicates a larger hierarchy and
further tuning will be necessary.

We further comment on tadpoles for the field strength $F_{56}$
localized at the fixed points in 6D models.  Such terms can directly
contribute to the Higgs mass term and receive quadratically divergent
corrections already at one loop~\cite{vonGersdorff:2002us}.  In our
previous work~\cite{Kawamura:2025}, we argued that a reflection
symmetry~\cite{GHU-CG&M,Scrucca:2003ut} can forbid the direct
contributions to Higgs mass terms from the tadpoles in our model.  We
have found, however, that the absence of direct zero-mode mass terms
is not by itself a complete argument for the protection of the Higgs
mass, as the tadpoles can mix the original zero modes with massive
Kaluza--Klein modes.  In the present analysis, we therefore assume
that the tadpole contributions are sufficiently small.  This
assumption amounts to a tuning of the corresponding localized terms,
in addition to the tuning discussed above.  Under this assumption, the
vacuum remains close to the Wilson line direction, and the
Wilson-line-based analysis is applicable.

The organization of this paper is as follows.  In section~2, we review
the SU(6) gauge-Higgs unification model with rank-reducing discrete
BCs on $T^2/\Z4$, emphasizing the ingredients used in the SU(9)
construction.  In section~3, we construct the SU(9) grand unified
model and analyze the zero modes of gauge and matter fields.  In
section~4, we discuss anomaly cancellation, gauge coupling relations,
proton stability and the EWSB.  Section~5 is devoted to conclusions
and discussions.  A detailed analysis of the zero modes is given in
Appendix~\ref{app:SU4}.

%%%%%%%%%%%%%%%%%%%%%%%%%%%%%%%%%%%%%%%%%%%%%%%%%%%%%%%%%%%%%%%%
%
\section{SU(6) gauge-Higgs unification model}
\label{sec:SU6GH}
%
%%%%%%%%%%%%%%%%%%%%%%%%%%%%%%%%%%%%%%%%%%%%%%%%%%%%%%%%%%%%%%%%
In this section, we review the SU(6) model in
ref.~\cite{Kawamura:2025} in some detail.  As seen in the subsequent
sections, several ingredients and results in the SU(6) model are
directly used to examine the SU(9) model. For more comprehensive
discussions, please refer to ref.~\cite{Kawamura:2025}.

%%%%%%%%%%%%%%%%%%%%%%%%%%%%%%%%%%%%%%%%%%%%%%%%%%%%
\subsection{Orbifold $T^2/{\mathbb Z}_4$}
\label{sec:T2Z4}
%%%%%%%%%%%%%%%%%%%%%%%%%%%%%%%%%%%%%%%%%%%%%%%%%%%%
 
We suppose a 6D space-time $M^4 \times T^2/{\mathbb Z}_4$, where $M^4$
is the 4D Minkowski space-time and $T^2/{\mathbb Z}_4$ is a
two-dimensional (2D) orbifold. Coordinates on $M^4$ and
$T^2/{\mathbb Z}_4$ are denoted by $x^\mu$ ($\mu=0,1,2,3$) (or $x$ for
short) and $z=x^5+ix^6$, respectively.  The orbifold
$T^2/{\mathbb Z}_4$ is given by imposing the identification under the
translation ${\cal T}_1:z\to z+1$ and the $\pi/2$ rotation
${\cal R}_0:z\to i z$ on the system :
$z \sim {\cal T}_1z\sim {\cal R}_0z$. Here, for simplicity, we take
the unit length of the torus lattice $T^2$ as $2\pi R =1$, where $R$
is regarded as a radius.  For the $\pi/2$ rotation,
${\cal R}_0^4={\cal I}$ holds, where ${\cal I}$ is the identity
operation.\footnote{ For the case of an action on fermion fields,
  ${\cal R}_0^4=-{\cal I}$ is generally allowed since ${\cal R}_0^4$
  corresponds to the $2\pi$ spatial rotation.}  Using ${\cal R}_0$ and
${\cal T}_{1}$, we can define the translations along $i^{m-1}$
$(m=1,\dots,4)$ direction as
\begin{align}
{\cal T}_{m}={\cal R}_0^{m-1}{\cal T}_1{\cal R}_0^{1-m},
  \qquad {\cal T}_{m}:z\to z+i^{m-1}.
\label{2D-Tn}
\end{align}
Since ${\cal T}_{m}$ are translations, they commute with each other
and obey $[{\cal T}_{m},{\cal T}_{m'}]=0$ for any pair $(m,m')$.  In
addition, as $i$ is the fourth root of unity, these translations
satisfy the relations:
\begin{align}\label{ztrans-rels}
\prod_{m=1}^4{\cal T}_m={\cal I}, \qquad {\cal T}_1{\cal T}_3
={\cal T}_2{\cal T}_4={\cal I}.
\end{align}
These relations are summarized as
$({\cal T}_m {\cal R}_0^p)^{4/p} = {\cal I}$ with $p=1, 2$~\cite{S&S}.

The translations and the $\pi/2$ rotations generally induce
non-trivial twists in the representation space of the symmetry group,
and the Lagrangian is invariant under these twists.  The twist
matrices corresponding to these operations are denoted by the Italic
character symbols, e.g., $R_0$ for ${\cal R}_0$ and $T_m$ for
${\cal T}_m$, which are unitary matrices belonging to the fundamental
representation.  These matrices are constrained by the relations
satisfied by the corresponding translations and rotations. For
example, the relations
${\cal T}_{m}={\cal R}_0^{m-1}{\cal T}_1{\cal R}_0^{1-m}$ and
$[{\cal T}_m,{\cal T}_{m'}]=0$ give the constraints
$T_{m}=R_0^{m-1}T_1R_0^{1-m}$ and $[T_m,T_{m'}]=0$ for the twist
matrices. All the twist matrices discussed below satisfy these
constraints.

%%%%%%%%%%%%%%%%%%%%%%%%%%%%%%%%%%%%%%%%%%%%%%%%%%%%
\subsection{Twist matrices, boundary conditions and Wilson line phases}
\label{sec:twist_su6}
%%%%%%%%%%%%%%%%%%%%%%%%%%%%%%%%%%%%%%%%%%%%%%%%%%%%

The twist matrices of the SU(6) model are given by, 
\bequ
 % R_0=\left(\begin{array}{cc|cccc}
 R_0=\left(\begin{array}{ccVcccc}
                -1&&&&&\\
                 &1&&&&\\ \hline
                 &&i&&&\\
                 &&&-1&&\\
                 &&&&-i&\\
                 &&&&&1
              \end{array} \right), \qquad
 % T_1=\left(\begin{array}{cc|cccc}
 T_1=\left(\begin{array}{ccVcccc}
                 &1&&&&\\
                 1&&&&&\\ \hline
                 &&1&&&\\
                 &&&1&&\\
                 &&&&1&\\
                 &&&&&1
              \end{array} \right), 
\label{Eq:SU(6)-w/WL}
\eequ
by which the SU(6) symmetry is broken down to ${\rm U}(1)^4$.
We note that the latter matrix is a U(6) element and assume 
the theory has a global U(6) symmetry.
With these twist matrices, we have both discrete and continuous Wilson
line phases, corresponding to the upper-left $2\times 2$ and
lower-right $4\times 4$ subblocks of the matrices, respectively. The
upper-left $2\times 2$ submatrices cannot be simultaneously
diagonalized by unitary and gauge transformations. Hence, the rank
reduction inevitably occurs with these twist matrices. In addition,
depending on the values of continuous Wilson line phases, the
remaining gauge symmetry can change from ${\rm U}(1)^4$ at a vacuum.

The twist matrices in eq.~\eqref{Eq:SU(6)-w/WL} constrain the BCs on
bulk fields as follows.  Let $\Phi(x,z,\bar{z})$ be a bulk 6D field
which forms an SU(6) multiplet belonging to an irreducible
representation ${\bf R}$ of SU(6).  We denote the corresponding twist
matrices in a general representation ${\bf R}$, including the
fundamental representation, by $R_0^{(\bf R )}$ and $T_1^{(\bf R )}$.
The BCs are given by
\begin{align}
  \Phi(x,z+1,\bar{z}+1)=\eta_T T_1^{(\bf R )}\Phi(x,z,\bar{z}), \qquad 
  \Phi(x,iz,-i\bar{z})=\eta_R R_0^{(\bf R )}\Phi(x,z,\bar{z}), 
  \label{Eq:bcgen1}
\end{align}
where $\eta_T$ and $\eta_R$ are some phase factors defined for each
field and satisfy the relations $\eta_T^2 = 1$ and $\eta_R^4 = 1$
derived from $T_1 T_3 = I$ and $R_0^4 = I$.

These BCs admit zero modes for the extra-dimensional components of the
gauge fields realizing the d.o.f. of the continuous Wilson line
phases.  Here, the continuous Wilson line phases are non-integrable
phases of, for instance,
$W_1=e^{ig(\langle A_z \rangle + \langle A_{\bar{z}} \rangle)}T_1$
with the gauge coupling $g$ and VEV of the zero modes in
$A_z = (A_5 - iA_6)/2$ and
$A_{\bar{z}} = (A_5 + iA_6)/2(= A_{z}^\dag)$ where $A_5$ and $A_6$ are
the gauge fields along the coordinates $x^5$ and $x^6$,
respectively. Under the ${\mathbb Z}_4$ rotation, the VEVs
$\langle A_z \rangle$ and $\langle A_{\bar{z}} \rangle$ change as
\begin{align}
R_0 \langle A_z \rangle R_0^{-1} = i \langle A_z \rangle,
\qquad
R_0 \langle A_{\bar{z}} \rangle R_0^{-1} = -i \langle A_{\bar{z}} \rangle.
\label{Gtr-<Az>su6}
\end{align}

There are tree-level potentials for the zero modes of the gauge
fields.  The VEVs of the zero modes along the
flat direction of the potential can be simplified by gauge
transformations of U(1)$^4$ and are parametrized by a complex
parameter $\alpha$ as
\bequ
% g\VEV{A_z}=2\pi\left(\begin{array}{cc|cccc}
g\VEV{A_z}=2\pi\left(\begin{array}{ccVcccc}
  0&&&&&\\
   &0&&&&\\ \hline
   &&&&&\alpha\\
   &&\alpha&&&\\
   &&&\alpha&&\\
   &&&&\alpha&
              \end{array} \right)
 =g\VEV{A_{\bar z}}^\dagger,
 \label{Eq:<Az>su6}
\eequ
where $2\pi R=1$ is taken.

The complex parameter $\alpha$ in eq.~\eqref{Eq:<Az>su6} is
parametrized by two real parameters $a$ and $b$ as
$\alpha = (a-ib)/2$.  Here, we note that the parameter $a$ ($b$) does
not correspond to $\VEV{A_5}$ ($\VEV{A_6}$), since the matrix
explicitly written in eq.~\eqref{Eq:<Az>su6} is not hermitian.  For
the twist matrices fixed as in eq.~\eqref{Eq:SU(6)-w/WL}, the
unbroken low-energy symmetries depend on the values of $a$ and $b$,
whereas integer shifts, sign flips and the exchange of $a$ and $b$ do
not affect the low-energy physics.  As stated, the twist matrices in
eq.~\eqref{Eq:SU(6)-w/WL} break ${\rm SU}(6)$ to ${\rm U}(1)^4$, which
is realized around $(a,b)=(0,0)$ in the field space of the continuous
Wilson line phases. On the other hand, around the specific point
$(a,b)=(0,1/2)$, the electroweak symmetry and its breaking can be
realized. To see this, let us perform a basis change by a unitary
transformation, after which $R_0$ and $W_1$ for $(a,b)=(0,1/2)$ take
\bequ
% R_0=\left(\begin{array}{cc:cc|cc}
R_0=\left(\begin{array}{ccDccVcc}
  -1&&&&&\\
    &-1&&&&\\ \hdashline
    &&1&&&\\
    &&&1&&\\ \hline
    &&&&i&\\
    &&&&&-i
              \end{array} \right), \qquad
 % W_1=\left(\begin{array}{cc:cc|cc}
 W_1=\left(\begin{array}{ccDccVcc}
                 &&1&&&\\
                 &&&1&&\\ \hdashline
                 1&&&&&\\
                 &1&&&&\\ \hline
                 &&&&&1\\
                 &&&&1&
              \end{array} \right). 
\label{Eq:SU(6)GHU-R0W1}
\eequ
The bulk ${\rm SU}(6)$ symmetry is broken down to ${\rm SU}(2)\sub{I}\times {\rm SU}(2)\sub{II}\times {\rm U}(1)_y\times {\rm U}(1)_A\times {\rm U}(1)\sub{III}$ by $R_0$, where the generators of the three ${\rm U}(1)$'s in the fundamental representation are respectively defined as
\begin{align}
\label{Eq:SU(6)GHU-U(1)generators_pre}
  T_y&=\frac1{2\sqrt6} t_y, && t_y={\rm diag}(1,1,1,1,-2,-2), \\
  T_A&=\frac1{2\sqrt2} t_A, && t_A={\rm diag}(1,1,-1,-1,0,0), \\
  T_{\rm III}&=\frac1{2} t_{\rm III}, && t_{\rm III}={\rm diag}(0,0,0,0,1,-1).
\label{Eq:SU(6)GHU-U(1)generators}
\end{align}
Hereafter, we take the normalization of a U(1)$_x$ generator $T_x$ in
the fundamental representation as ${\rm Tr}(T_x)^2=1/2$. In addition,
we refer to eigenvalues of $t_x$ as U(1)$_x$ charges.  Due to the
action of $W_1$ in eq.~\eqref{Eq:SU(6)GHU-R0W1}, the exchange of the
two SU(2) groups,
${\rm SU}(2)\sub{I}\leftrightarrow {\rm SU}(2)\sub{II}$ occurs, and
the signs of the charges of U(1)$_A$ and ${\rm U}(1)\sub{III}$ flip.
Then, ${\rm SU}(2)\sub{I}\times {\rm SU}(2)\sub{II}$ symmetry is
broken down to its diagonal subgroup ${\rm SU}(2)_{\rm D}$, and
${\rm U}(1)_A\times {\rm U}(1)\sub{III}$ is broken down to nothing.
The resultant symmetry is ${\rm SU}(2)_{\rm D}\times {\rm U}(1)_y$,
which is further broken down to ${\rm U}(1)$ if the vacuum is away
from $(a,b)=(0,1/2)$.  We can identify this symmetry-breaking pattern
with the EWSB, in which the gauge and Higgs fields are unified.

%%%%%%%%%%%%%%%%%%%%%%%%%%%%%%%%%%%%%%%%%%%%%%%%%%%%
\subsection{Zero modes of bulk fields}
\label{sec:su6spectrum}
%%%%%%%%%%%%%%%%%%%%%%%%%%%%%%%%%%%%%%%%%%%%%%%%%%%%

In this SU(6) gauge-Higgs unification model, two Higgs doublets appear
as the zero modes of the extra-dimensional components of the gauge
field.  In addition, the quarks in each generation can be unified into
one multiplet without exotic quarks, as the zero modes of a bulk field
belonging to the \repr{15} representation of SU(6).

To see these zero modes, we discuss decompositions of the bulk
${\rm SU}(6)$ multiplet into representations of
${\rm SU}(2)\sub{I}\times {\rm SU}(2)\sub{II}\times {\rm U}(1)_y\times
{\rm U}(1)_A\times {\rm U}(1)\sub{III}$
and examine the BCs of the gauge
field. The fundamental representation \repr6 of ${\rm SU}(6)$
group is given as
\beqn
\repr6&=&\Ls\MyDecom{2}{1}{1}{1}{0}{2},\MyDecom{1}{2}{1}{-1}{0}{0},
\MyDecom{1}{1}{-2}{0}{1}{1},\MyDecom{1}{1}{-2}{0}{-1}{-1} \Rs^T,
\label{Eq:decom6}
\eeqn
where representations/charges of
$\Ls \rm{SU(2)}\sub{I},\rm{SU(2)}\sub{II}\Rs_{{\rm U(1)}_y,{\rm U(1)}_A,{\rm
    U(1)}\sub{III}}^{R_0}$ are shown at the indicated positions, and
$T$ in the right-hand side denotes the transpose. These U(1) charges are
identified with eigenvalues of $t_x$ in
eqs.~\eqref{Eq:SU(6)GHU-U(1)generators_pre}--\eqref{Eq:SU(6)GHU-U(1)generators}. The
$R_0$ charges correspond to ones of
$t_R={\rm diag}(-2,-2,0,0,1,3)\sim {\rm diag}(2,2,0,0,1,-1)$,
where $R_0=e^{2\pi i t_R/4}=i^{t_R}$.
Since $R_0$ charges are defined modulo 4, we represent them in the range $[-1,2]$
in what follows.

The adjoint representation $\repr{35}$ of SU(6) is contained in
$\bm 6\times \repr{\overline{6}}$
and is written as
\begin{align}
\label{Eq:SU(6)GHU-adjoint}  
  \repr{35}
  = \left(
  % \begin{array}{c:c|cc}
  \begin{array}{cDcVcc}
    \MyDecom{3}{1}{0}{0}{0}{0}&\MyDecom{2}{2}{0}{2}{0}{2}
    &\MyDecom{2}{1}{3}{1}{-1}{1}&\MyDecom{2}{1}{3}{1}{1}{-1}\\ \hdashline
                                &\MyDecom{1}{3}{0}{0}{0}{0}
    &\MyDecom{1}{2}{3}{-1}{-1}{-1}&\MyDecom{1}{2}{3}{-1}{1}{1}\\ \hline
                                &&&\MyDecom{1}{1}{0}{0}{2}{2}\\
                                &&&\vphantom{0}
  \end{array}
  \right)
  +3\times \MyDecom{1}{1}{0}{0}{0}{0},
\end{align}
where the matrix form is expressed in the same basis as in 
  eq.~\eqref{Eq:decom6}.
The lower-left entries, which are given by the complex conjugates of the upper-right ones, are suppressed. 
The diagonal entries, including the three singlets,
correspond to the adjoint representations of
${\rm SU}(2)\sub{I}\times {\rm SU}(2)\sub{II}\times {\rm U}(1)_y\times
{\rm U}(1)_A\times {\rm U}(1)\sub{III}$.

For $(a,b)=(0,1/2)$, zero modes are invariant under both BCs in
eq.~\eqref{Eq:bcgen1}, in which the twist matrices of the $\pi/2$
rotation and the translation are given by $R_0$ and $W_1$ in
eq.~\eqref{Eq:SU(6)GHU-R0W1}, respectively.  One finds that the action
of $R_0$ and $W_1$ in eq.~\eqref{Eq:SU(6)GHU-R0W1} on the fundamental
representation $\repr6$ is given by
\begin{align}
R_0\cdot \repr6&=\Ls-\MyDecom{2}{1}{1}{1}{0}{2},\MyDecom{1}{2}{1}{-1}{0}{0},
i\MyDecom{1}{1}{-2}{0}{1}{1},-i\MyDecom{1}{1}{-2}{0}{-1}{-1} \Rs^T,  \\
W_1\cdot \repr6&=\Ls\MyDecom{1}{2}{1}{-1}{0}{0},\MyDecom{2}{1}{1}{1}{0}{2},
\MyDecom{1}{1}{-2}{0}{-1}{-1},\MyDecom{1}{1}{-2}{0}{1}{1} \Rs^T.
\label{6translaw}
\end{align}
As stated, one sees that $W_1$ in eq.~\eqref{Eq:SU(6)GHU-R0W1}
exchanges ${\rm SU}(2)\sub{I}$ and ${\rm SU}(2)\sub{II}$ and flips the
sign of the charges of $ \rm{U}(1)_A$ and ${\rm U}(1)\sub{III}$.

For $A_\mu$, $\eta_R=1$ must be taken. Thus, the components that
have zero $R_0$ charge in eq.~\eqref{Eq:SU(6)GHU-adjoint} may admit
zero modes. We can show that, besides the singlet corresponding to $T_y$,
the linear combination $\MyDecom{3}{1}{0}{0}{0}{0}+ \MyDecom{1}{3}{0}{0}{0}{0}$
contains the zero mode, which realizes ${\rm SU}(2)_{\rm D}(\times {\rm U}(1)_y)$ gauge symmetry in the 4D effective theory.
We can understand it from the BCs of the $SU(2)_{\rm I}$ gauge boson $A_{\mu}^{({\rm I})}$
and the $SU(2)_{\rm II}$ gauge boson $A_{\mu}^{({\rm II})}$ given by
\beqn
&&A_{\mu}^{({\rm I})}(x, iz, -i\bar{z}) = A_{\mu}^{({\rm I})}(x, z, \bar{z}),\qquad
A_{\mu}^{({\rm II})}(x, iz, -i\bar{z}) = A_{\mu}^{({\rm II})}(x, z, \bar{z}), 
\label{Eq:SU(6)Amu-R0}\\
&&A_{\mu}^{({\rm I})}(x, z+1, \bar{z}+1) = A_{\mu}^{({\rm II})}(x, z, \bar{z}),\qquad
A_{\mu}^{({\rm II})}(x, z+1, \bar{z}+1) = A_{\mu}^{({\rm I})}(x, z, \bar{z}), 
\label{Eq:SU(6)Amu-W1}
\eeqn
under the $\Z 4$ rotation and the translation represented by the twist matrices
in \Eqref{Eq:SU(6)GHU-R0W1}.
They are rewritten as
\beqn
A_{\mu}^{(\pm)}(x, iz, -i\bar{z}) = A_{\mu}^{(\pm)}(x, z, \bar{z}),\qquad
A_{\mu}^{(\pm)}(x, z+1, \bar{z}+1) = \pm A_{\mu}^{(\pm)}(x, z, \bar{z}),
\label{Eq:SU(6)Amu(pm)-BC}
\eeqn
where $A_{\mu}^{(\pm)} \equiv (A_{\mu}^{({\rm I})} \pm A_{\mu}^{({\rm II})})/\sqrt{2}$.
Thus, $A_{\mu}^{(+)}$ contains the zero mode, namely the 
${\rm SU}(2)_{\rm D}$ gauge boson~\cite{gGHU,Kojima:2023mew}.

For $A_z$ $(=A_{\bar z}^\dag)$, $\eta_R=-i$ should hold. Thus, the
components whose $R_0$ charges are $1$ in
eq.~\eqref{Eq:SU(6)GHU-adjoint} may admit zero modes.  In the adjoint
representation \repr{35}, there are four such components, namely,
$\MyDecom{2}{1}{3}{1}{-1}{1}$, $\MyDecom{1}{2}{3}{-1}{1}{1}$,
$\Ls \MyDecom{2}{1}{3}{1}{1}{-1}\Rs^*$ and
$\Ls\MyDecom{1}{2}{3}{-1}{-1}{-1}\Rs^*$.
As in the case of the
$\rm{SU(2)_I}$ and $\rm{SU(2)_{II}}$ gauge bosons, the translation
exchanges $\MyDecom{2}{1}{3}{1}{-1}{1}$ and
$\MyDecom{1}{2}{3}{-1}{1}{1}$ ($\Ls \MyDecom{2}{1}{3}{1}{1}{-1}\Rs^*$
and $\Ls\MyDecom{1}{2}{3}{-1}{-1}{-1}\Rs^*$).
We see that the following two linear combinations
\beqn
  &&\MyDecom{2}{1}{3}{1}{-1}{1} + \MyDecom{1}{2}{3}{-1}{1}{1}, 
\qquad 
\Ls \MyDecom{2}{1}{3}{1}{1}{-1} + \MyDecom{1}{2}{3}{-1}{-1}{-1}\Rs^*,
\label{2hdmzeromode}
\eeqn 
have the zero modes. These zero modes are identified with
two Higgs doublets.

From the ${\rm U}(1)_y$ charges of the Higgs doublets, we can choose the
normalization of the weak hypercharge $Q_Y$ as\footnote{
For later convenience, the sign in the definition of $Q_Y$ is reversed compared to that in ref.~\cite{Kawamura:2025}.}
\begin{align}
Q_Y=-{2\sqrt6\over 6} T_y=-{1\over 6}t_y={\rm diag}(-1/6,-1/6,-1/6,-1/6,1/3,1/3).
\end{align}
We find $g_Y=\sqrt{3/2}g_6$ and $g_w=g_6/\sqrt{2}$, where $g_6$, $g_Y$
and $g_w$ are the effective 4D gauge coupling constant of the bulk
${\rm SU}(6)$, the ${\rm U}(1)$ weak hypercharge and the ${\rm SU}(2)$
gauge couplings in the SM gauge symmetry, respectively. A large value
of the Weinberg angle $\sin\theta_W=\sqrt{3}/2\simeq 0.87$ around the
compactification scale is predicted. To obtain a realistic low-energy
effective theory, effects from boundary operators, renormalization
groups and/or mixing with an additional ${\rm U}(1)$ gauge symmetry
must be included.

Let us discuss fermion fields.
A Weyl fermion in 6D is regarded as a 4D Dirac fermion or a pair of 4D Weyl fermions with opposite chiralities in 4D. A 6D Dirac fermion $\Psi$ consists of two 6D Weyl fermions such that
\beqn
\Psi_{+} \equiv \frac{1+\Gamma_7}{2}\Psi = \left(\begin{array}{c}
\psi_{(+,{\rm L})} \\
\psi_{(+,{\rm R})}
\end{array}\right),~~
\Psi_{-} \equiv \frac{1-\Gamma_7}{2}\Psi = \left(\begin{array}{c}
\psi_{(-,{\rm R})} \\
\psi_{(-,{\rm L})}
\end{array}\right),
\label{Eq:6DPsi}
\eeqn
where $\Psi_{+}$ and $\Psi_{-}$ are fermions with positive and negative 6D chirality, respectively,
$\Gamma_7$ is the chirality operator in 6D, and the subscripts $\pm$ and L(R) stand for the chiralities in 6D and 4D, respectively.
From the $\Z4$ invariance of the kinetic term and the transformation property of the covariant derivatives $D_z \to -iR_0D_zR_0^{-1}$ and $D_{\bar{z}} \to iR_0D_{\bar{z}}R_0^{-1}$, we have the relations among $\eta_R$ in eq.~\eqref{Eq:bcgen1}:
\beqn
\eta_{R}^{(+,{\rm R})} = i \eta_{R}^{(+,{\rm L})},~~
\eta_{R}^{(-,{\rm R})} = -i \eta_{R}^{(-,{\rm L})},
\label{Eq:etaR}
\eeqn
where $\eta_{R}^{(\pm,{\rm R})}$ and $\eta_{R}^{(\pm,{\rm L})}$ are $\eta_R$ of $\psi_{(\pm,{\rm R})}$ and $\psi_{(\pm,{\rm L})}$, respectively~\cite{G&K,GK&M}.

In BCs for fermion fields, as long as the constraints in
eq.~\eqref{Eq:etaR} are satisfied, there is a freedom to choose
$\eta_R$ for the $\Z4$ rotation, in addition to $R_0$ in
eq.~\eqref{Eq:bcgen1}, for each multiplet. (We have similar d.o.f. for
the translation, $\eta_T$.)  We can choose $\eta_R$ to make any
components having the same $R_0$ charge in a multiplet invariant under
the $\Z4$ rotation.  For example, $\eta_R=-1=e^{2\pi i\cdot 2/4}$
effectively shifts $R_0$ charges by 2, and then, the component
$\MyDecom{2}{1}{1}{1}{0}{2}$ in eq.~\eqref{Eq:decom6} is $\Z4$
invariant. Since this component is not an eigenstate of $W_1$, we see
that there is no zero mode.  It is understood that, since the twist
matrices in eq.~\eqref{Eq:SU(6)GHU-R0W1} satisfy $R_0W_1=-W_1R_0$,
each operation of $W_1$ flips the sign of $R_0$ eigenvalues and shifts
$R_0$ charges by 2 for $\MyDecom{2}{1}{1}{1}{0}{2}$,
$\MyDecom{1}{2}{1}{-1}{0}{0}$, $\MyDecom{1}{1}{-2}{0}{1}{1}$ and
$\MyDecom{1}{1}{-2}{0}{-1}{-1}$ which belong to $\repr6$ of SU(6) in
eq.~\eqref{Eq:decom6}. This implies that we need matter fields
belonging to even-rank tensor multiplets of SU(6) to obtain zero
modes.

In the SU(6) model, quarks in the SM in each generation appear as the
zero modes of a Weyl fermion belonging to the $\repr{15}$
representation, which is the second rank anti-symmetric tensor and is
decomposed as
\begin{align}
  \repr{15}
   =
   % \left(\begin{array}{c:c|cc}
   \left(\begin{array}{cDcVcc}
                \MyDecom{1}{1}{2}{2}{0}{0}&\MyDecom{2}{2}{2}{0}{0}{2}
                   &\MyDecom{2}{1}{-1}{1}{1}{-1}&\MyDecom{2}{1}{-1}{1}{-1}{1}\\ \hdashline
                 &\MyDecom{1}{1}{2}{-2}{0}{0}
                   &\MyDecom{1}{2}{-1}{-1}{1}{1}&\MyDecom{1}{2}{-1}{-1}{-1}{-1}\\ \hline
                 &&&\MyDecom{1}{1}{-4}{0}{0}{0}\\
                 &&&\vphantom{0}
   \end{array} \right).
\label{Eq:SU(6)GHU-rank2}
\end{align}
To clarify zero modes contained in fermion fields belonging to
$\bm{15}$, as an example, let us suppose $\eta_R=1$. Then, the
components $\MyDecom{1}{1}{2}{2}{0}{0}$, $\MyDecom{1}{1}{2}{-2}{0}{0}$
and $\MyDecom{1}{1}{-4}{0}{0}{0}$ may admit zero modes.  The
translation with $W_1$ exchanges $\MyDecom{1}{1}{2}{2}{0}{0}$ and
$\MyDecom{1}{1}{2}{-2}{0}{0}$.  Hence, the corresponding fields,
denoted by $\psi^{(1)}$ and $\psi^{(2)}$, obey the BCs as
\beqn
\hspace{-1cm} && \psi^{(1)}(x, iz, -i\bar{z}) = \psi^{(1)}(x, z,
\bar{z}),\qquad \psi^{(2)}(x, iz, -i\bar{z}) = \psi^{(2)}(x, z,
\bar{z}),
\label{Eq:SU(6)psi-R0}\\
\hspace{-1cm}
&& \psi^{(1)}(x, z+1, \bar{z}+1) = \eta_T \psi^{(2)}(x, z, \bar{z}),\qquad
\psi^{(2)}(x, z+1, \bar{z}+1) = \eta_T \psi^{(1)}(x, z, \bar{z}), 
\label{Eq:SU(6)psi-W1}
\eeqn
under the $\Z 4$ rotation and the translation $z \to z+1$. 
They are rewritten as
\beqn
\psi^{(\pm)}(x, iz, -i\bar{z}) = \psi^{(\pm)}(x, z, \bar{z}),\qquad
\psi^{(\pm)}(x, z+1, \bar{z}+1) = \pm \eta_T \psi^{(\pm)}(x, z, \bar{z}),
\label{Eq:SU(6)psi(pm)-BC}
\eeqn where
$\psi^{(\pm)} \equiv (\psi^{(1)} \pm \psi^{(2)})/\sqrt{2}$.  Thus, for
the cases with $\eta_T=\pm 1$, $\psi^{(\pm)}$ contains a zero mode,
which is an ${\rm SU(2)}_{\rm D}$ singlet having $Q_Y=-1/3$. On the
other hand, $\MyDecom{1}{1}{-4}{0}{0}{0}$ is odd under the action of
$W_1$.\footnote{ We note that $\MyDecom{1}{1}{-4}{0}{0}{0}$ is
  understood as the anti-symmetric tensor product of
  $\MyDecom{1}{1}{-2}{0}{1}{1}$ and $\MyDecom{1}{1}{-2}{0}{-1}{-1}$ in
  eq.~\eqref{Eq:decom6}, which are exchanged under the action of $W_1$
  in eq.~\eqref{Eq:SU(6)GHU-R0W1}.}  The corresponding field obeys the
same BCs as $\psi^{(-)}$. Thus, if we choose $\eta_T=-1$, the
component $\MyDecom{1}{1}{-4}{0}{0}{0}$ has the zero mode, which is an
${\rm SU(2)}_{\rm D}$ singlet having $Q_Y=2/3$.  In a similar way, we
can find zero modes with different $\eta_R$.  When we take $\eta_R=i$
$(-i)$, a linear combination
$\MyDecom{2}{1}{-1}{1}{1}{-1}\pm \MyDecom{1}{2}{-1}{-1}{-1}{-1}$
($\MyDecom{2}{1}{-1}{1}{-1}{1}\pm \MyDecom{1}{2}{-1}{-1}{1}{1}$)
contains a zero mode for $\eta_T=\pm 1$.  This zero mode is an
${\rm SU(2)}_{\rm D}$ doublet having $Q_Y=1/6$.  When we take
$\eta_R=-1$, the component $\MyDecom{2}{2}{2}{0}{0}{2}$ may admit zero
modes.  The four d.o.f. in $\MyDecom{2}{2}{2}{0}{0}{2}$ can be
expressed as $\phi_{13}$, $\phi_{14}$, $\phi_{23}$ and $\phi_{24}$,
where the subscripts correspond to their positions in the matrix form
in eq.~\eqref{Eq:SU(6)GHU-rank2}. Under the action of $W_1$, they
transform as $\phi_{13}\to \phi_{31}=-\phi_{13}$,
$\phi_{14}\to \phi_{32}=-\phi_{23}$,
$\phi_{23}\to\phi_{41}=-\phi_{14}$ and
$\phi_{24}\to \phi_{42}=-\phi_{24}$.  Thus, if we choose $\eta_T=1$,
the component $\phi_{14}-\phi_{23}$ has the zero mode, which is an
${\rm SU(2)}_{\rm D}$ singlet and has $Q_Y=-1/3$. On the other hand,
if we choose $\eta_T=-1$, the rest three components have the zero
modes, which form an ${\rm SU(2)}_{\rm D}$ triplet and have $Q_Y=-1/3$.

From the above, one sees that the quarks in a generation, namely the
quark singlets $U_{\rm R}$ and $D_{\rm R}$ and the quark doublet
$Q_{\rm L}$ in the SM, where the subscript R (L) implies a
right-handed (left-handed) 4D Weyl fermion, can appear as the zero modes
from a 6D Weyl fermion of $\bm{15}$ representation.  For $\Psi_+$ with
$\eta_T=-1$ and $(\eta_R^{(+,{\rm L})},\eta_R^{(+,{\rm R})})=(-i,1)$,
the component
$ \MyDecom{2}{1}{-1}{1}{-1}{1}- \MyDecom{1}{2}{-1}{-1}{1}{1}$ in
$\psi_{(+,{\rm L})}$ gives $Q_{\rm L}$, and the components
$\MyDecom{1}{1}{-4}{0}{0}{0}$ and
$\MyDecom{1}{1}{2}{2}{0}{0}-\MyDecom{1}{1}{2}{-2}{0}{0}$ in
$\psi_{(+,{\rm R})}$ respectively give $U_{\rm R}$ and
$D_{\rm R}$.  For $\Psi_-$ with $\eta_T=-1$ and
$(\eta_R^{(-,{\rm L})},\eta_R^{(-,{\rm R})})=(i,1)$, the component
$\MyDecom{2}{1}{-1}{1}{1}{-1}- \MyDecom{1}{2}{-1}{-1}{-1}{-1}$
in $\psi_{(-,{\rm L})}$ gives $Q_{\rm L}$, and as
in the $\Psi_+$ case,
the same components in $\psi_{(-,{\rm R})}$, $\MyDecom{1}{1}{-4}{0}{0}{0}$ and
$\MyDecom{1}{1}{2}{2}{0}{0}-\MyDecom{1}{1}{2}{-2}{0}{0}$, give $U_{\rm R}$ and
$D_{\rm R}$.  We summarize how the quarks appear as the zero modes of
a 6D Weyl fermion of $\bm{15}$ representation with $\eta_T=-1$ as
follows:
\beqn
\Psi_{+}(\repr{15}) =\begin{cases} \psi_{(+,{\rm
        R})}(\repr{15}) \ni \begin{cases}
        \MyDecom{1}{1}{2}{2}{0}{0}-\MyDecom{1}{1}{2}{-2}{0}{0}
          \Rightarrow D_{\rm R} \\
          \MyDecom{1}{1}{-4}{0}{0}{0} \Rightarrow U_{\rm R}
 \end{cases} \\
  \psi_{(+,{\rm L})}(\repr{15}) \ni
\MyDecom{2}{1}{-1}{1}{-1}{1}- \MyDecom{1}{2}{-1}{-1}{1}{1}
\Rightarrow Q_{\rm L},
\end{cases}
\label{Eq:quarksinSU(6)p}
\eeqn
or
\beqn
\Psi_{-}(\repr{15}) =\begin{cases}
  \psi_{(-,{\rm R})}(\repr{15}) \ni \begin{cases}
\MyDecom{1}{1}{2}{2}{0}{0}-\MyDecom{1}{1}{2}{-2}{0}{0}
\Rightarrow D_{\rm R} \\
  \MyDecom{1}{1}{-4}{0}{0}{0} \Rightarrow U_{\rm R}
 \end{cases} \\
  \psi_{(-,{\rm L})}(\repr{15}) \ni
\MyDecom{2}{1}{-1}{1}{1}{-1}- \MyDecom{1}{2}{-1}{-1}{-1}{-1}
\Rightarrow Q_{\rm L}.
\end{cases}
\label{Eq:quarksinSU(6)n}
\eeqn

Depending on the 6D chirality of the Weyl fermions, up and down-type Higgs doublets ($H_u$ and $H_d$)
are identified with components in eq.~\eqref{2hdmzeromode}. 
For the case with $\Psi_+$, 
$\MyDecom{2}{1}{3}{1}{-1}{1} + \MyDecom{1}{2}{3}{-1}{1}{1}$
and
$\Ls \MyDecom{2}{1}{3}{1}{1}{-1} + \MyDecom{1}{2}{3}{-1}{-1}{-1}\Rs^*$
are identified with $H_u^*$ and $H_d^*$, respectively. 
For the case with $\Psi_-$, 
$\MyDecom{2}{1}{3}{1}{-1}{1} + \MyDecom{1}{2}{3}{-1}{1}{1}$
and
$\Ls \MyDecom{2}{1}{3}{1}{1}{-1} + \MyDecom{1}{2}{3}{-1}{-1}{-1}\Rs^*$
are identified with $H_d$ and $H_u$, respectively.

The second-rank tensors do not contain components with the same
quantum numbers as the leptons in the SM. In addition, odd-rank
tensors, such as third-rank ones, do not admit zero modes in the SU(6)
model. To incorporate leptons, bulk fermions in higher-rank tensor
representations are required, unless the SM fermions are introduced as
brane fields. Another option is to modify the ${\rm U(1)}_Y$ operator
with the help of additional ${\rm U(1)}$ gauge symmetries. As shown in
the following sections, such a modification of the ${\rm U(1)}_Y$
operator compared to the SU(6) case is naturally realized in an SU(9)
model, and the leptons, namely, the lepton doublet $(L_{\rm L})$, the
charged lepton $(E_{\rm R})$ and the right-handed neutrino
$(N_{\rm R})$, arise as zero modes from a second-rank tensor of SU(9).

%%%%%%%%%%%%%%%%%%%%%%%%%%%%%%%%%%%%%%%%%%%%%%%%%%%%
\subsection{Vacuum structure}
\label{sec:su6vac}
%%%%%%%%%%%%%%%%%%%%%%%%%%%%%%%%%%%%%%%%%%%%%%%%%%%%
Let us discuss the vacuum structure of the SU(6) model. If the vacuum deviates from the specific point characterized by
eq.~\eqref{Eq:SU(6)GHU-R0W1}, namely
$(a,b)=(\delta a ,1/2+\delta b)\neq (0,1/2)$ with nonvanishing
$\delta a$ and $\delta b$, the EWSB occurs via the Hosotani
mechanism. The fluctuations $\delta a$ and $\delta b$ are identified
with some of the components of the two Higgs doublets $H_u$ and $H_d$.  If
suitable bulk fields are incorporated, the EWSB vacuum can be obtained
as a global minimum of the effective potential for the Wilson line
phases.

The one-loop contributions to the effective potential generated by
bulk fields are derived in ref.~\cite{Kawamura:2025}. For later
convenience, we summarize the result. As explained below
eq.~\eqref{Eq:<Az>su6}, the Wilson line phases are parametrized by two
real parameters $a$ and $b$.  Let ${\cal V}_{\bf R}^{[\beta_T]}(a,b)$
be a contribution to the effective potential from a real bosonic
d.o.f. of a bulk field belonging to an irreducible representation
$\bf{R}$.  The parameter $\beta_T$ is defined by
$\eta_T=e^{2\pi{ i \beta_T}}$, and can have $0$ or $1/2$.  Let us
define \bequ {\cal V}^{[\beta_T]}(q_1,q_2)=
-\frac1{16\pi^7R^4}\sum_{(w_1,w_2)\neq(0,0)} \frac{\cos(2\pi(w_1
  (q_1+\beta_T)+w_2(q_2+\beta_T)))}{(w_1^2+w_2^2)^3},
\label{Eq:V-quartet}
\eequ where $w_1$ and $w_2$ are integers. In this expression, the
torus size $2\pi R$ is explicitly shown, and the divergent constant
term is removed.  Using it, the contributions to the effective
potential ${\cal V}_{\bf R}^{[\beta_T]}(a,b)$ are expressed as
follows: \beqn {\cal V}^{[\beta_T]}_\repr6(a,b)&=&{\cal
  V}^{[\beta_T]}(a,b),
\label{Eq:V-6} \\
{\cal V}^{[\beta_T]}_\repr{15}(a,b)&=&{\cal V}^{[\beta_T]}(a+b,a-b)+{\cal V}^{[\beta_T]}(a,b+1/2)+{\cal V}^{[\beta_T]}(a+1/2,b), 
\label{Eq:V-15} \\\notag
{\cal V}^{[\beta_T]}_\repr{21}(a,b)&=&{\cal V}^{[\beta_T]}(2a,2b)+{\cal V}^{[\beta_T]}(a+b,a-b)
\\ && +{\cal V}^{[\beta_T]}(a,b+1/2)+{\cal V}^{[\beta_T]}(a+1/2,b), 
\label{Eq:V-21} \\\notag
{\cal V}^{[\beta_T]}_\repr{35}(a,b)&=&{\cal V}^{[\beta_T]}(2a,2b)+2{\cal V}^{[\beta_T]}(a+b,a-b)
\\ && +2{\cal V}^{[\beta_T]}(a,b+1/2)+2{\cal V}^{[\beta_T]}(a+1/2,b), 
\label{Eq:V-35}  \\ 
{\cal V}^{[\beta_T]}_\repr{56}(a,b)&=&{\cal V}^{[\beta_T]}(3a,3b)+{\cal V}^{[\beta_T]}(2a+b,a-2b)+{\cal V}^{[\beta_T]}(2a-b,a+2b) \nn\\
                       &&+{\cal V}^{[\beta_T]}(2a,2b+1/2)+{\cal V}^{[\beta_T]}(2a+1/2,2b) \nn\\
                       &&+{\cal V}^{[\beta_T]}(a+b,a-b+1/2)+{\cal V}^{[\beta_T]}(a+b+1/2,a-b) \nn\\
                       &&+4{\cal V}^{[\beta_T]}(a,b)+{\cal V}^{[\beta_T]}(a+1/2,b+1/2), 
\label{Eq:V-56} \\
{\cal V}^{[\beta_T]}_\repr{70}(a,b)&=&{\cal V}^{[\beta_T]}(2a+b,a-2b)+{\cal V}^{[\beta_T]}(2a-b,a+2b) \nn\\
                       &&+{\cal V}^{[\beta_T]}(2a,2b+1/2)+{\cal V}^{[\beta_T]}(2a+1/2,2b) \nn\\
                       &&+2{\cal V}^{[\beta_T]}(a+b,a-b+1/2)+2{\cal V}^{[\beta_T]}(a+b+1/2,a-b) \nn\\
                       &&+5{\cal V}^{[\beta_T]}(a,b)+2{\cal V}^{[\beta_T]}(a+1/2,b+1/2), 
\label{Eq:V-70} \\
{\cal V}^{[\beta_T]}_\repr{20}(a,b)&=&{\cal V}^{[\beta_T]}(a+b,a-b+1/2)+{\cal V}^{[\beta_T]}(a+b+1/2,a-b) \nn\\
                       &&+{\cal V}^{[\beta_T]}(a,b)+{\cal V}^{[\beta_T]}(a+1/2,b+1/2).
\label{Eq:V-20} 
\eeqn
These contributions are independent of a choice of the parameter $\eta_R$ for each bulk matter field, as well as the difference between representations $\bf{R}$ and $\repr{\overline{R}}$.
One can see that these contributions are invariant under integer shifts, sign flips and exchange of $a$ and $b$.

Using the above, for a given bulk matter contents, we can obtain the
one-loop effective potential for the Wilson line phases $a$ and $b$.
The vacuum structure of the potential is numerically examined.  We
have shown in ref.~\cite{Kawamura:2025} that there exist global minima that are slightly displaced
from $(a,b)=(0,1/2)$ with suitable bulk matter contents.
Such global minima are regarded as the EWSB vacuum. The smallness of the
deviation from $(a,b)=(0,1/2)$ means a hierarchy between the scales of
a compactification and the EWSB.

In this analysis, for simplicity, we restrict our analysis to vacua
along the flat direction of the tree-level potentials for the zero
modes of the gauge fields.  In a more realistic setup, however,
provided that the doublet VEVs are sufficiently small compared with
the compactification scale, the vacuum need not be exactly aligned
with a tree-level flat direction.  At the tree level, the quadratic
terms vanish, whereas the quartic terms come from the commutator term
in the field strength and have the same form as the SUSY D-term
contributions. The quadratic terms are generated via loop effects, and
thus their naive scale is suppressed by the loop factor relative to
the compactification scale~\cite{Kojima:2008ky}.  For suitable
loop-induced quadratic terms, Higgs zero modes corresponding to
non-flat directions can acquire small VEVs.\footnote{ In a realistic
  setup, a complete calculation of the one-loop effective potential is
  generally difficult. It is therefore reasonable to analyze the
  vacuum instead within the low-energy effective
  theory~\cite{Haba:2005kc}.}  As discussed in the next subsection,
the quadratic terms can also receive contributions from the localized
tadpoles. Such contributions are generally sensitive to the cutoff
scale and are assumed to be sufficiently small in our model.

%%%%%%%%%%%%%%%%%%%%%%%%%%%%%%%%%%%%%%%%%%%%%%%%%%%%
\subsection{Effects from localized tadpoles}
\label{sec:su6tad}
%%%%%%%%%%%%%%%%%%%%%%%%%%%%%%%%%%%%%%%%%%%%%%%%%%%%

We finally comment on the effects of localized tadpoles in the SU(6)
model.  If the unbroken gauge group at an orbifold fixed point
contains a U(1) factor, the localized operator linear in the
corresponding U(1) component of the field strength $F_{56}$ is
generically allowed and can be generated
radiatively~\cite{vonGersdorff:2002us}. Their coefficients may contain
both bare (tree-level) localized contributions and loop corrections.

In the SU(6) model studied in ref.~\cite{Kawamura:2025}, we argued
that a reflection symmetry~\cite{GHU-CG&M,Scrucca:2003ut} forbids
direct localized-tadpole contributions to the zero-mode Higgs mass
terms, which would otherwise introduce cutoff-sensitive contributions
into the Higgs potential. This argument is incomplete, because
localized tadpoles can generate not only bilinear terms involving the
Higgs zero modes but also mixing between the original zero modes and
massive Kaluza--Klein modes, and the reflection symmetry does not in
general forbid the mixing terms. Such mixing can modify the classical
background and the full Kaluza--Klein mass matrix, and may affect the
stability of the flat direction.  Indeed, even if one neglects such
mixing and truncates the analysis to the original zero-mode sector,
ref.~\cite{Hasegawa:2015vqa} shows that the tadpole term can drive the
vacuum away from the tree-level flat direction.  Therefore, the
absence of a direct localized mass term for the Higgs zero modes is
not sufficient to justify an analysis restricted to fluctuations
around tree-level flat directions.  Consequently, the analysis in
ref.~\cite{Kawamura:2025} does not establish the stability of the
hierarchy between the electroweak and compactification scales against
localized tadpoles. In the absence of an additional suppression
mechanism, maintaining this hierarchy may require a fine-tuning of the
localized tadpole coefficients to obtain a cancellation between
loop-induced contributions and bare localized counterterms. In the
following discussion, we assume that the effects of localized tadpoles
are sufficiently small that the Wilson-line background and the
Kaluza--Klein spectrum used below provide a valid approximation. A
detailed analysis of tadpole-induced deformations is left for future
work.

%%%%%%%%%%%%%%%%%%%%%%%%%%%%%%%%%%%%%%%%%%%%%%%%%%%%%%%%%%%%%%%%
%
\section{SU(9) grand unified model}
\label{sec:SU9GUT}
%
%%%%%%%%%%%%%%%%%%%%%%%%%%%%%%%%%%%%%%%%%%%%%%%%%%%%%%%%%%%%%%%%

The SU(6) model reviewed in the previous section has several
attractive features.  First, the rank-reducing discrete BCs on
$T^2/\mathbb Z_4$ realize the symmetry structure relevant to
electroweak gauge-Higgs unification.  Second, two Higgs doublets arise
from the zero modes of the extra-dimensional components of the gauge
field.  Third, the quarks in each generation can be obtained from a
bulk fermion in the $\bm{15}$ representation without exotic zero
modes.  The electroweak symmetry can then be broken by the Hosotani
mechanism through the dynamics of the continuous Wilson line phases.

At the same time, the SU(6) model is not a grand unified model of the
SM gauge interactions: the color interaction is not included in
the gauge group.  In addition, leptons are not naturally
obtained as zero modes and an excessively large value of
    the Weinberg angle at the compactification scale is predicted, 
in the minimal setup.  These points motivate an extension in which the
SU(6) structure is embedded into a larger gauge group containing the
color SU(3) and an additional U(1) symmetry.

A natural possibility is SU(9), which contains
${\rm SU(6)}\times {\rm SU(3)}\times {\rm U(1)}$ as a subgroup.  As
will be shown, the positive features of the SU(6) model are inherited,
while the unsatisfactory aspects are improved by the SU(9) grand
unified model constructed in this section.  In particular, the SU(6)
sector responsible for the Wilson line phases and the Higgs doublets
can be retained, while the additional SU(3) factor is identified with
color.  The extra Abelian generator also plays an essential role in
embedding the weak hypercharge and in obtaining leptons as zero modes
from SU(9) multiplets without including exotic matter.  Furthermore,
after redefining the weak hypercharge by taking the extra U(1) into
account, the Weinberg angle takes a value close to the observed value,
as will be seen in the next section.

%%%%%%%%%%%%%%%%%%%%%%%%%%%%%%%%%%%%%%%%%%%%%%%%%%%% 
\subsection{Boundary conditions on $T^2/\Z4$}
\label{sec:BCs}
%%%%%%%%%%%%%%%%%%%%%%%%%%%%%%%%%%%%%%%%%%%%%%%%%%%%

We consider a model based on an SU(9) gauge symmetry, by adding the color SU(3) factor to the SU(6) model, with
the following BCs on 6D bulk fields,
\begin{align}
  R_0=\left(
  % \begin{array}{cc|ccccIccc}
  \begin{array}{ccVccccIccc}
    -1&&&&&&&&\\
    &1&&&&&&&\\ \hline
    &&i&&&&&&\\
    &&&-1&&&&&\\
    &&&&-i&&&&\\
    &&&&&1&&&\\ \whline
    &&&&&&-i&&\\
    &&&&&&&-i&\\
    &&&&&&&&-i
  \end{array}\right), \qquad
  T_1=\left(
  % \begin{array}{cc|ccccIccc}
  \begin{array}{ccVccccIccc}
    &1&&&&&&&\\
    1&&&&&&&&\\ \hline
    &&1&&&&&&\\
    &&&1&&&&&\\
    &&&&1&&&&\\
    &&&&&1&&&\\ \whline
    &&&&&&1&&\\
    &&&&&&&1&\\
    &&&&&&&&1
  \end{array}\right),
\label{eq:SU(9)-w/WL}
\end{align}
where $R_0$ and $T_1$ are elements of U(9) and we assume the
  theory has a global U(9) symmetry.  These BCs contain the twist
matrices in eq.~\eqref{Eq:SU(6)-w/WL} as the upper-left $6\times 6$
submatrices.  With these BCs, the SU(9) symmetry is broken down to
U(1)$^4 \times$SU(4).  We note that the rank is reduced from 8 to 7 by
the upper-left block of $T_1$.

The VEV of the extra-dimensional components of the gauge fields can be parameterized, with the normalization of $2\pi R=1$, as
\bequ
% g\VEV{A_z}=2\pi\left(\begin{array}{cc|ccccIccc}
g\VEV{A_z}=2\pi\left(\begin{array}{ccVccccIccc}
                       0&&&&&&&&\\
                        &0&&&&&&&\\ \hline
                        &&&&&\alpha&&&\\
                        &&\alpha&&&&&&\\
                        &&&\alpha&&&&&\\
                        &&&&\alpha&&&&\\ \whline
                        &&&&&&0&&\\
                        &&&&&&&0&\\
                        &&&&&&&&0
              \end{array} \right)
 =g\VEV{A_{\bar z}}^\dagger, 
\label{Eq:<Az>}
\eequ 
where $\alpha = (a-ib)/2$ ($a$, $b$ : real parameters),
referring to \Eqref{Eq:<Az>su6}.

Taking a basis change by a unitary transformation and rearranging
the rows and columns in the twist matrices, $R_0$ and $W_1\equiv e^{ig(\langle A_z \rangle + \langle A_{\bar{z}} \rangle)}T_1$ with $(a,b)=(0,1/2)$ are given by
\bequ
 % R_0=\left(\begin{array}{cc:cc|ccIccc}
 R_0=\left(\begin{array}{ccDccVccIccc}
                -1&&&&&&&&\\
                 &-1&&&&&&&\\ \hdashline
                 &&1&&&&&&\\
                 &&&1&&&&&\\ \hline
                 &&&&i&&&&\\
                 &&&&&-i&&&\\ \whline
                 &&&&&&-i&&\\
                 &&&&&&&-i&\\
                 &&&&&&&&-i
              \end{array} \right),
\label{Eq:SU(9)R0-permute}
\eequ
and
\bequ
 % W_1=\left(\begin{array}{cc:cc|ccIccc}
 W_1=\left(\begin{array}{ccDccVccIccc}
                 &&1&&&&&&\\
                 &&&1&&&&&\\ \hdashline
                 1&&&&&&&&\\
                 &1&&&&&&&\\ \hline
                 &&&&&1&&&\\
                 &&&&1&&&&\\ \whline
                 &&&&&&1&&\\
                 &&&&&&&1&\\
                 &&&&&&&&1
              \end{array} \right). 
\label{Eq:SU(9)W1-permute}
\eequ
The bulk SU(9) symmetry is broken down to 
${\rm SU}(2)\sub{I}\times {\rm SU}(2)\sub{II}\times {\rm SU}(4)\times {\rm U}(1)_A
\times {\rm U}(1)_B\times {\rm U}(1)_C$ by $R_0$, where the SU(4) symmetry is broken down to SU(3) by $W_1$. 
Here, the generators of the three U(1)'s are respectively defined as
\begin{align}
  T_A&=\frac1{2\sqrt2} t_A, && t_A={\rm diag}(1,1,-1,-1,0,0,0,0,0), 
\label{Eq:SU(9)TA}\\
  T_B&=\frac1{2\sqrt{10}} t_B, && t_B={\rm diag}(1,1,1,1,-4,0,0,0,0), 
\label{Eq:SU(9)TB}\\
  T_C&=\frac1{6\sqrt{10}} t_C, && t_C={\rm diag}(4,4,4,4,4,-5,-5,-5,-5).
\label{Eq:SU(9)GHU-U(1)generators}
\end{align}
For later convenience, we define the generator $T_D$ of a U(1) subgroup of SU(4) as
\begin{align}
  T_D&=\frac1{2\sqrt{6}} t_D, && t_D={\rm diag}(0,0,0,0,0,-3,1,1,1).
\label{Eq:SU(9)GHU-U(1)D}
\end{align}
As in the SU(6) case, the generators $T_x$ are normalized 
as ${\rm Tr}(T_x)^2 = \frac{1}{2}$ for the fundamental representation of SU(9), and 
we refer to eigenvalues of $t_x$ as U(1)$_x$ charges. Some indices are permuted by $W_1$, and this permutation is roughly expressed by the exchange of the two SU(2) groups, ${\rm SU}(2)\sub{I}\leftrightarrow {\rm SU}(2)\sub{II}$, 
and the sign flips of the charges of U(1)$_A$ and U(1)$_{\rm III}$ defined below. Then, due to the action of $W_1$, ${\rm SU}(2)\sub{I}\times {\rm SU}(2)\sub{II}$ symmetry is broken down to its diagonal subgroup ${\rm SU}(2)_{\rm D}$~\cite{Kawamura:2025} 
and ${\rm SU}(4)\times {\rm U}(1)_A\times {\rm U}(1)_B\times {\rm U}(1)_C$ symmetry is broken down to 
${\rm SU}(3)\times {\rm U}(1)_y\times {\rm U}(1)_{\rm IV}$.
Here, the generators of ${\rm U}(1)_y$ and ${\rm U}(1)_{\rm IV}$ are respectively defined as
\begin{align}
  T_y&=\frac1{2\sqrt6} t_y, && t_y={\rm diag}(1,1,1,1,-2,-2,0,0,0), 
\label{Eq:SU(9)Ty}\\
  T_{\rm IV}&=\frac{1}{6} t_{\rm IV}, && t_{\rm IV}={\rm diag}(1,1,1,1,1,1,-2,-2,-2).
\label{Eq:SU(9)GHU-U(1)generators2}
\end{align}
We note that the SM hypercharge gauge group ${\rm U}(1)_{Y}$ is not
the ${\rm U}(1)_{y}$ itself but a subgroup of
${\rm U}(1)_{y}\times{\rm U}(1)_{\rm IV}$, where ${\rm U}(1)_{\rm IV}$
is defined by
${\rm SU}(9)\supset {\rm SU}(6)\times {\rm SU}(3) \times {\rm
  U}(1)_{\rm IV}$.  On the breaking from
${\rm SU}(4)\times {\rm U}(1)_A\times {\rm U}(1)_B\times {\rm U}(1)_C$
to ${\rm SU}(3)\times {\rm U}(1)_y\times {\rm U}(1)_{\rm IV}$, the
broken symmetries generated by the Cartan subalgebra are
${\rm U}(1)_A\times {\rm U}(1)_{\rm III}$, where the generator of
${\rm U}(1)_{\rm III}$ is defined as
\begin{align}
  T_{\rm III}&=\frac1{2} t_{\rm III}, && t_{\rm III}={\rm diag}(0,0,0,0,1,-1,0,0,0).
\label{Eq:SU(9)GHU-U(1)generators-U1III}
\end{align}
The relations among the generators of U(1)'s are expressed as
\begin{align}
t_{y}= \frac{3}{5} t_B + \frac{1}{10} t_C + \frac{1}{2} t_D,~~~
t_{\rm III}= -\frac{1}{5} t_B + \frac{1}{20} t_C + \frac{1}{4} t_D,~~~
t_{\rm IV}= \frac{1}{4} t_C - \frac{3}{4} t_D.
\label{Eq:SU(9)GHU-RelsU(1)}
\end{align}

At the specific point $(a,b)=(0,1/2)$, the gauge symmetry is enhanced to ${\rm SU}(2)_{\rm D}\times {\rm SU}(3) \times {\rm U}(1)_{y}\times {\rm U}(1)_{\rm IV}$. Away from this point, it is broken down to ${\rm SU}(3)\times {\rm U}(1)^2$. This symmetry-breaking pattern can be considered as the EWSB via the Hosotani mechanism, if SU(2)$_{\rm D}$ is regarded as SU(2)$_{\rm L}$ in the SM and the surviving ${\rm U}(1)^2$ symmetry contains the electromagnetic symmetry denoted by U(1)$_{\rm em}$. In this case, the Higgs field should appear as a zero mode of $A_z$ and $A_{\bar z}$, that is, be unified with the gauge field. An extra U(1) symmetry is supposed to be broken down around the compactification scale by the VEV of some extra scalar field. In the following sections, we examine this type of the grand unified and gauge-Higgs unification model.

More precisely, we consider the following symmetry-breaking pattern:
\beqn
{\rm SU}(9) 
&\xrightarrow{(a,b)=(0,1/2)}& 
{\rm SU}(2)_{\rm D}\times {\rm SU}(3)_{\rm C}\times {\rm U}(1)_{y}\times {\rm U}(1)_{\rm IV}
\nonumber \\
&\xrightarrow{\rm singlet~VEV}&
{\rm SU}(2)_{\rm D}\times {\rm SU}(3)_{\rm C}\times {\rm U}(1)_Y 
\nonumber \\
&\xrightarrow{\rm Hosotani~mechanism}&
{\rm U}(1)_{\rm em} \times {\rm SU}(3)_{\rm C},
\label{Eq:brpattern}
\eeqn where SU(3) is denoted as SU(3)$_{\rm C}$ by considering it as
the gauge group describing the strong interaction in the SM, singlet
VEV stands for the VEV of a scalar field which is a singlet under
${\rm SU}(2)_{\rm D}\times {\rm SU}(3)_{\rm C}\times {\rm U}(1)_Y$,
and U(1)$_Y$ is the gauge group relating to the weak hypercharge in
the SM. Here, the generator of U(1)$_Y$
is defined by a linear combination of the generators of U(1)$_{y}$ and U(1)$_{\rm IV}$, as shown in the next subsection.

%%%%%%%%%%%%%%%%%%%%%%%%%%%%%%%%%%%%%%%%%%%%%%%%%%%%
\subsection{Zero modes of bulk fields}
\label{sec:ZeroModes}
%%%%%%%%%%%%%%%%%%%%%%%%%%%%%%%%%%%%%%%%%%%%%%%%%%%%

Let us see how the SM particles come from the zero modes of specific
bulk fields.  We postpone the discussion of anomaly-free matter
contents to the next section.

We use the subgroup
${\rm SU}(2)\sub{I}\times {\rm SU}(2)\sub{II}\times {\rm SU}(3)\times
{\rm U}(1)_y\times {\rm U}(1)_A\times {\rm U}(1)\sub{III}\times {\rm
  U}(1)\sub{IV}$ of SU(9) instead of
${\rm SU}(2)\sub{I}\times {\rm SU}(2)\sub{II}\times {\rm SU}(4)\times
{\rm U}(1)_A \times {\rm U}(1)_B\times {\rm U}(1)_C$, because it is
easy to understand how they originate from and are identified with the
SM particles.\footnote{ The study on zero modes based on the subgroup
  ${\rm SU}(2)\sub{I}\times {\rm SU}(2)\sub{II}\times {\rm
    SU}(4)\times {\rm U}(1)_A \times {\rm U}(1)_B\times {\rm U}(1)_C$
  of SU(9) is provided in Appendix~\ref{app:SU4}.} Here,
${\rm SU}(2)\sub{I}\times {\rm SU}(2)\sub{II}\times {\rm U}(1)_y\times
{\rm U}(1)_A\times {\rm U}(1)\sub{III}$ is the subgroup of SU(6) in
${\rm SU}(9) \supset {\rm SU}(6) \times {\rm SU}(3) \times {\rm
  U}(1)_{\rm IV}$.

The action of the $\Z4$ rotation is embedded in U(9), and we can define charges corresponding to $R_0$ by the generator 
$t_R=(2,2,0,0,1,-1,-1,-1,-1)$, 
where $R_0=e^{2\pi i t_R/4}=i^{t_R}$. 
We note that the following relation holds
\beqn
t_R=
    \frac13t_y+t_A+t_{\rm III} + \frac{1}{9}(5t_{\rm IV}+t_{\rm E}),
\label{Eq:tR-relation-SU3}
\eeqn
where $t_{\rm E}$ is the U(1) charge in the subgroup ${\rm SU}(9)\times {\rm U}(1)$ of U(9) defined by
\beqn
  t_{\rm E}={\rm diag}(1,1,1,1,1,1,1,1,1),
\label{Eq:U(1)generator}
\eeqn
and it is not a gauge generator, but rather for bookkeeping purposes.

As seen in the previous section, in the SU(6) model, odd-rank tensors of SU(6) do not admit zero modes, and the quarks in each generation can originate from the anti-symmetric rank-2 tensors of SU(6), that is, the bulk fermions in the $\repr{15}$ representation of SU(6). Hence, we study zero modes by paying attention to even-rank tensors of SU(6), which is constructed from the fundamental representation \repr9 of SU(9) group and/or its conjugate $\repr{\overline{9}}$.

For reference, \repr9 of SU(9) group and its conjugate $\repr{\overline{9}}$ are decomposed into
\beqn
 \repr{9}&=&(\repr{6}, \repr1)_{1} + (\repr1, \repr3)_{-2},
\label{Eq:decom9(6+3)-SU3}\\
 \repr{\overline{9}}&=&(\repr{\overline{6}}, \repr1)_{-1} + (\repr1, \repr{\overline{3}})_{2},
\label{Eq:decom9bar(6+3)-SU3}
\eeqn
respectively, under ${\rm SU}(6) \times {\rm SU}(3) \times {\rm U}(1)_{\rm IV}$,
and they are further decomposed into
\beqn
\hspace{-1cm}
 \repr9&=&\Ls\MyDecomC{2}{1}{1}{1}{1}{0}{1}{2},\MyDecomC{1}{2}{1}{1}{-1}{0}{1}{0},
                    \MyDecomC{1}{1}{1}{-2}{0}{1}{1}{1},\MyDecomC{1}{1}{1}{-2}{0}{-1}{1}{-1},\MyDecomC{1}{1}{3}{0}{0}{0}{-2}{-1} \Rs^T,
\label{Eq:decom9-SU3}\\
\hspace{-1cm}
 \repr{\overline{9}}&=&\Ls\MyDecomC{2}{1}{1}{-1}{-1}{0}{-1}{2},\MyDecomC{1}{2}{1}{-1}{1}{0}{-1}{0},
                           \MyDecomC{1}{1}{1}{2}{0}{-1}{-1}{-1},\MyDecomC{1}{1}{1}{2}{0}{1}{-1}{1},\MyDecomC{1}{1}{\overline{3}}{0}{0}{0}{2}{1}\Rs,
\label{Eq:decom6bar-SU3}
\eeqn
where representations/charges of 
$\Ls {\rm SU}(2)\sub{I},{\rm SU}(2)\sub{II}, {\rm SU}(3)\Rs_{{\rm U}(1)_y,{\rm U}(1)_A,{\rm U}(1)\sub{III}, {\rm U}(1)\sub{IV}}^{R_0}$ are shown at the indicated positions, and $T$ in \Eqref{Eq:decom9-SU3} denotes the transpose.

First, we examine the anti-symmetric rank-3 tensors: $\repr{84}$,
which is decomposed into
\beqn
 \repr{84}&=&(\repr{20}, \repr1)_{3} + (\repr{15}, \repr3)_{0} 
+ (\repr6, \repr{\overline{3}})_{-3}
 + (\repr1, \repr1)_{-6},
\label{Eq:decom84-SU3}
\eeqn
under ${\rm SU}(6) \times {\rm SU}(3) \times {\rm U}(1)_{\rm IV}$.
Since $(\repr{20}, \repr1)_{3}$ and $(\repr6, \repr{\overline{3}})_{-3}$ are
odd-rank tensor multiplets of SU(6),
no zero modes appear from these multiplets.
On the other hand,
$(\repr{15}, \repr3)_{0}$
can have zero modes. It
is decomposed into
\bequ
  (\repr{15}, \repr3)_{0}
   % = \left(\begin{array}{c:c|cc}
   = \left(\begin{array}{cDcVcc}
                \MyDecomC{1}{1}{3}{2}{2}{0}{0}{-1}&\MyDecomC{2}{2}{3}{2}{0}{0}{0}{1}
                   &\MyDecomC{2}{1}{3}{-1}{1}{1}{0}{2}&\MyDecomC{2}{1}{3}{-1}{1}{-1}{0}{0}\\ \hdashline
                 &\MyDecomC{1}{1}{3}{2}{-2}{0}{0}{-1}
                   &\MyDecomC{1}{2}{3}{-1}{-1}{1}{0}{0}&\MyDecomC{1}{2}{3}{-1}{-1}{-1}{0}{2}\\ \hline
                 &&0&\MyDecomC{1}{1}{3}{-4}{0}{0}{0}{-1}\\
                 &&&0
              \end{array} \right),
\label{Eq:SU(6)GHU-rank3-SU3}
\eequ
where 
the suppressed lower-left triangle is given from the upper-right triangle so that it becomes anti-symmetric. 
We note that this expression is given from eq.~\eqref{Eq:SU(6)GHU-rank2} 
by adding ${\bf 3}$ (0) as the SU(3) representation (U(1)$_{\rm IV}$ charge) 
and shifting the $R_0$ charges by $-1$. 
As seen from the observation in subsection \ref{sec:su6spectrum}, quarks can be obtained as zero modes of $(\repr{15}, \repr3)_{0}$ with $\eta_T = -1$, 
based on the definition of the weak  hypercharge $Q_{Y}$ as
\beqn
Q_Y &\equiv& -\frac{1}{6}t_y +ct_{\rm IV}, 
\label{Eq:hyperchargeInSU9c}
\eeqn where the coefficient $c$ in front of $t_{\rm IV}$ will later be
determined by the assignment of leptons.  Since the
${\rm U}(1)\sub{IV}$ charge of $(\repr{15}, \repr3)_{0}$ is zero,
there is no contribution to the multiplet from the term related to
$t_{\rm IV}$.  When $\repr{84}$ has $\eta_T = -1$, the multiplet
$(\repr1, \repr1)_{-6}$ in \Eqref{Eq:decom84-SU3} does not contain a
zero mode, leading to no exotics from this quark multiplet.

Next, we investigate the anti-symmetric rank-2 tensors of SU(9): $\repr{36}$,
which is decomposed into
\beqn
 \repr{36}&=&(\repr{15}, \repr1)_{2} + (\repr6, \repr3)_{-1}
 + (\repr1, \repr{\overline{3}})_{-4},
\label{Eq:decom36-SU3}
\eeqn
under ${\rm SU}(6) \times {\rm SU}(3) \times {\rm U}(1)_{\rm IV}$.
No zero modes stem from the multiplet $(\repr6, \repr3)_{-1}$, 
since $\repr6$ is the rank-1 tensor of SU(6), while
$(\repr{15}, \repr1)_{2}$
can have zero modes. The latter is decomposed into
\bequ
  (\repr{15}, \repr1)_{2}
   % = \left(\begin{array}{c:c|cc}
   = \left(\begin{array}{cDcVcc}
                \MyDecomC{1}{1}{1}{2}{2}{0}{2}{0}&\MyDecomC{2}{2}{1}{2}{0}{0}{2}{2}
                   &\MyDecomC{2}{1}{1}{-1}{1}{1}{2}{-1}&\MyDecomC{2}{1}{1}{-1}{1}{-1}{2}{1}\\ \hdashline
                 &\MyDecomC{1}{1}{1}{2}{-2}{0}{2}{0}
                   &\MyDecomC{1}{2}{1}{-1}{-1}{1}{2}{1}&\MyDecomC{1}{2}{1}{-1}{-1}{-1}{2}{-1}\\ \hline
                 &&0&\MyDecomC{1}{1}{1}{-4}{0}{0}{2}{0}\\
                 &&&0
              \end{array} \right),
\label{Eq:SU(6)GHU-rank2-SU3}
\eequ
in a similar way as \Eqref{Eq:SU(6)GHU-rank3-SU3}.  We recall
that the weak hypercharge of each lepton is $2/3$ less than that of
the corresponding quark and that the ${\rm U}(1)\sub{IV}$ charge of
$(\repr{15}, \repr1)_{2}$ is 2 greater than that of
$(\repr{15}, \repr3)_{0}$.  This means that, if we set the coefficient
$c$ in eq.~\eqref{Eq:hyperchargeInSU9c} to be $-1/3$, \beqn Q_Y
&\equiv& -\frac{1}{6}t_y - \frac{1}{3}t_{\rm IV} =
-\frac{2}{\sqrt{6}}T_y - 2T_{\rm IV}
\nonumber \\
&=&{\rm diag}(-1/2,-1/2,-1/2,-1/2,0,0,2/3,2/3,2/3),
\label{Eq:hypercharge-SU3}
\eeqn
the zero modes of $(\repr{15}, \repr1)_{2}$ with $\eta_T = -1$ can be identified with leptons. 
When $\repr{36}$ has $\eta_T = -1$, the multiplet $(\repr1, \repr{\overline{3}})_{-4}$ in \Eqref{Eq:decom36-SU3} does not contain a zero mode, 
  leading to no exotics from this lepton multiplet.
Using the weak hypercharge defined by \Eqref{Eq:hypercharge-SU3}, its normalization is determined as $T_Y=\sqrt{\frac{3}{14}} Q_Y$. 
Later, we use this relation in the analysis of the gauge coupling unification.

Here, we summarize how quarks and leptons come from in our model.
There are two types of assignments for quarks and leptons.
Quarks originate from the zero modes of the following bulk fermions with $\eta_T = -1$:
\beqn
\Psi_{+}(\repr{84}) =\begin{cases}
 \psi_{(+,{\rm R})}(\repr{84}) \ni \begin{cases}
  \MyDecomC{1}{1}{3}{2}{2}{0}{0}{-1} - \MyDecomC{1}{1}{3}{2}{-2}{0}{0}{-1}  \Rightarrow D_{\rm R} \\
  \MyDecomC{1}{1}{3}{-4}{0}{0}{0}{-1} \Rightarrow U_{\rm R}
 \end{cases} \\
 \psi_{(+,{\rm L})}(\repr{84}) \ni \MyDecomC{2}{1}{3}{-1}{1}{-1}{0}{0} - \MyDecomC{1}{2}{3}{-1}{-1}{1}{0}{0}
\Rightarrow Q_{\rm L},
\end{cases}
\label{Eq:quarks+SU3}
\eeqn
or
\beqn
\Psi_{-}(\repr{84}) =\begin{cases}
 \psi_{(-,{\rm R})}(\repr{84}) \ni \begin{cases}
  \MyDecomC{1}{1}{3}{2}{2}{0}{0}{-1} -\MyDecomC{1}{1}{3}{2}{-2}{0}{0}{-1} \Rightarrow D_{\rm R} \\
  \MyDecomC{1}{1}{3}{-4}{0}{0}{0}{-1} \Rightarrow U_{\rm R}
 \end{cases} \\
 \psi_{(-,{\rm L})}(\repr{84}) \ni\MyDecomC{2}{1}{3}{-1}{1}{1}{0}{2} - \MyDecomC{1}{2}{3}{-1}{-1}{-1}{0}{2}
\Rightarrow Q_{\rm L} .
\end{cases}
\label{Eq:quarks-SU3}
\eeqn
In a similar way, leptons originate from the zero modes of the following bulk fermions with $\eta_T = -1$:
\beqn
\Psi_{+}(\repr{36}) =\begin{cases}
 \psi_{(+,{\rm R})}(\repr{36}) \ni \begin{cases}
  \MyDecomC{1}{1}{1}{2}{2}{0}{2}{0} - \MyDecomC{1}{1}{1}{2}{-2}{0}{2}{0} \Rightarrow E_{\rm R} \\
  \MyDecomC{1}{1}{1}{-4}{0}{0}{2}{0} \Rightarrow N_{\rm R}
 \end{cases} \\
 \psi_{(+,{\rm L})}(\repr{36}) \ni \MyDecomC{2}{1}{1}{-1}{1}{-1}{2}{1} - \MyDecomC{1}{2}{1}{-1}{-1}{1}{2}{1}
\Rightarrow L_{\rm L} ,
\end{cases}
\label{Eq:leptons+SU3}
\eeqn
or
\beqn
\Psi_{-}(\repr{36}) =\begin{cases}
 \psi_{(-,{\rm R})}(\repr{36}) \ni \begin{cases}
  \MyDecomC{1}{1}{1}{2}{2}{0}{2}{0} - \MyDecomC{1}{1}{1}{2}{-2}{0}{2}{0} \Rightarrow E_{\rm R} \\
  \MyDecomC{1}{1}{1}{-4}{0}{0}{2}{0} \Rightarrow N_{\rm R}
 \end{cases} \\
  \psi_{(-,{\rm L})}(\repr{36}) \ni
  \MyDecomC{2}{1}{1}{-1}{1}{1}{2}{-1} - \MyDecomC{1}{2}{1}{-1}{-1}{-1}{2}{-1}
\Rightarrow L_{\rm L} .
\end{cases}
\label{Eq:leptons-SU3}
\eeqn
We note that eq.~\eqref{Eq:etaR} is used to realize zero modes and $D_{\rm R}$, $U_{\rm R}$, $E_{\rm R}$ and $N_{\rm R}$ denote 4D right-handed Weyl fermions, not their charge-conjugated left-handed fields.

Next, we study whether the SM gauge bosons and the SM Higgs doublet arise as the zero modes of the 6D SU(9) gauge boson or not. The 6D SU(9) gauge boson $(A_{\mu}, A_{z}, A_{\bar{z}})$ in the adjoint representation \repr{80} is decomposed into
\beqn
 \repr{80}&=&(\repr{35}, \repr1)_{0} + (\repr6, \repr{\overline{3}})_{3}
 + (\repr{\overline{6}}, \repr{3})_{-3} + (\repr{1}, \repr{8})_{0} + (\repr1, \repr1)_{0}
,
\label{Eq:decom80-SU3}
\eeqn
under ${\rm SU}(6) \times {\rm SU}(3) \times {\rm U}(1)_{\rm IV}$.
In \Eqref{Eq:decom80-SU3}, the gluon $G_{\mu}$ and an extra U(1) gauge boson $A_{\mu}^{\rm (IV)}$ appear from the zero modes of the bulk fields in $(\repr{1}, \repr{8})_{0}$ and $(\repr{1}, \repr{1})_{0}$, respectively, and no scalar fields appear from zero modes of the bulk fields with these representations.
Moreover, the bulk fields in $(\repr6, \repr{\overline{3}})_{3} + (\repr{\overline{6}}, \repr{3})_{-3}$ have no zero modes, because $\repr6$ and $\repr{\overline{6}}$ are odd-rank tensors of SU(6).
In eq.~\eqref{Eq:decom80-SU3}, $(\repr{35}, \repr1)_{0}$ is decomposed into
\beqn
 (\repr{35}, \repr1)_{0}
   % &=& \left(\begin{array}{c:c|cc}
   &=& \left(\begin{array}{cDcVcc}
                \MyDecomC{3}{1}{1}{0}{0}{0}{0}{0}&\MyDecomC{2}{2}{1}{0}{2}{0}{0}{2}
                   &\MyDecomC{2}{1}{1}{3}{1}{-1}{0}{1}&\MyDecomC{2}{1}{1}{3}{1}{1}{0}{-1}\\ \hdashline
                 &\MyDecomC{1}{3}{1}{0}{0}{0}{0}{0}
                   &\MyDecomC{1}{2}{1}{3}{-1}{-1}{0}{-1}&\MyDecomC{1}{2}{1}{3}{-1}{1}{0}{1}\\ \hline
                 &&\vphantom{0}&\MyDecomC{1}{1}{1}{0}{0}{2}{0}{2}\\
                 &&&\vphantom{0}
              \end{array} \right) 
\nonumber\\
 &~& +~ 3 \times \MyDecomC{1}{1}{1}{0}{0}{0}{0}{0},
\label{Eq:SU(6)GHU-adjoint-SU3}
\eeqn
where 
the suppressed lower-left triangle is given from the upper-right triangle so that it becomes hermitian.
We note that this expression is given from eq.~\eqref{Eq:SU(6)GHU-adjoint} 
by adding ${\bf 1}$ (0) as the SU(3) representation (U(1)$_{\rm IV}$ charge).

For $A_\mu$ in $(\repr{35}, \repr{1})_{0}$, $\eta_R=1$ must be taken, and we can show, besides the singlet corresponding to $T_y$, that the linear combination $\MyDecomC{3}{1}{1}{0}{0}{0}{0}{0}+ \MyDecomC{1}{3}{1}{0}{0}{0}{0}{0}$,
contains the zero mode, which  realizes ${\rm SU}(2)_{\rm D}(\times {\rm U}(1)_y)$ gauge symmetry in the 4D effective theory, in a similar way as explained in the subsection \ref{sec:su6spectrum}.
This zero mode is regarded as the SU(2)$_{\rm L}$ weak boson $W_{\mu}$.

Two of the four singlets, $(\repr1, \repr1)_{0}$ in
eq.~\eqref{Eq:decom80-SU3} and
$3 \times (\repr1, \repr1, \repr1)^{0}_{0,0,0,0}$ in
eq.~\eqref{Eq:SU(6)GHU-adjoint-SU3}, have zero modes, which realize
${\rm U}(1)_{y}\times {\rm U}(1)_{\rm IV}$ gauge symmetry.  The
remaining ${\rm U}(1)_y \times {\rm U}(1)_{\rm IV}$ gauge symmetry can
be expressed by ${\rm U}(1)_Y \times {\rm U}(1)_{Y^{\bot}}$ where
${\rm U}(1)_Y$ and ${\rm U}(1)_{Y^{\bot}}$ represent the U(1) gauge
group regarding the weak hypercharge and its orthogonal U(1) symmetry,
respectively. The generator of ${\rm U}(1)_{Y^{\bot}}$ is defined by
\beqn
Q_{Y^{\bot}} \equiv \sqrt{\frac{1}{7}}(\sqrt{6} T_y - T_{\rm
  IV}),
\label{Eq:QYbot-SU3}
\eeqn
with the normalization as the fundamental representation of SU(9).

For $A_z$ and $A_{\bar z}$, $\eta_R=-i$ and $\eta_R = i$ should hold, respectively, and we see that the following linear combinations
\beqn
\hspace{-1cm}
  &&\MyDecomC{2}{1}{1}{3}{1}{-1}{0}{1} + \MyDecomC{1}{2}{1}{3}{-1}{1}{0}{1},~~~~
\MyDecomC{2}{1}{1}{-3}{-1}{-1}{0}{1} + \MyDecomC{1}{2}{1}{-3}{1}{1}{0}{1}
~~~~{\rm for}~A_z,
\label{Eq:SU(6)GHU-Hu-SU3}\\
\hspace{-1cm}
  &&\MyDecomC{2}{1}{1}{3}{1}{1}{0}{-1} + \MyDecomC{1}{2}{1}{3}{-1}{-1}{0}{-1},~~~~
\MyDecomC{2}{1}{1}{-3}{-1}{1}{0}{-1} + \MyDecomC{1}{2}{1}{-3}{1}{-1}{0}{-1}
~~~~{\rm for}~A_{\bar z},
\label{Eq:SU(6)GHU-Hd-SU3}
\eeqn 
have the zero modes.
We note that the origin of Higgs doublets is essentially inherited from the SU(6) model and
$\MyDecomC{2}{1}{1}{3}{1}{-1}{0}{1} + \MyDecomC{1}{2}{1}{3}{-1}{1}{0}{1}$ ($\MyDecomC{2}{1}{1}{-3}{-1}{-1}{0}{1} + \MyDecomC{1}{2}{1}{-3}{1}{1}{0}{1}$) and $\MyDecomC{2}{1}{1}{-3}{-1}{1}{0}{-1} + \MyDecomC{1}{2}{1}{-3}{1}{-1}{0}{-1}$ ($\MyDecomC{2}{1}{1}{3}{1}{1}{0}{-1} + \MyDecomC{1}{2}{1}{3}{-1}{-1}{0}{-1}$) are particles and antiparticles.

Here, we summarize how the SM gauge bosons ($G_{\mu}$, $W_{\mu}$, $B_{\mu}$) and the two 
Higgs doublets ($H_u$, $H_d$) come from. They are generated as
\beqn
A_M =\begin{cases}
 A_{\mu} \ni \begin{cases}
  \MyDecomC{1}{1}{8}{0}{0}{0}{0}{0} \Rightarrow G_{\mu} \\
  \MyDecomC{3}{1}{1}{0}{0}{0}{0}{0}+\MyDecomC{1}{3}{1}{0}{0}{0}{0}{0} \Rightarrow W_{\mu} \\
  \MyDecomC{1}{1}{1}{0}{0}{0}{0}{0}+\MyDecomC{1}{1}{1}{0}{0}{0}{0}{0},
  \MyDecomC{1}{1}{1}{0}{0}{0}{0}{0} \Rightarrow A^{(y)}_{\mu}, A^{({\rm IV})}_{\mu}
     \sim B_{\mu}, B^{\bot}_{\mu}
 \end{cases} \\
 A_{z} \ni  \MyDecomC{2}{1}{1}{3}{1}{-1}{0}{1} + \MyDecomC{1}{2}{1}{3}{-1}{1}{0}{1}
\Rightarrow H_u^* \\
 A_{\bar{z}} \ni \MyDecomC{2}{1}{1}{3}{1}{1}{0}{-1} + \MyDecomC{1}{2}{1}{3}{-1}{-1}{0}{-1}
\Rightarrow H_d,
\end{cases}
\label{Eq:bosons-SU3}
\eeqn
where 
$A^{(y)}_{\mu}$ and $A^{({\rm IV})}_{\mu}$ are the gauge bosons of ${\rm U}(1)_{y}$ and ${\rm U}(1)_{\rm IV}$, respectively, the gauge bosons $B_{\mu}$ and $B^{\bot}_{\mu}$ associated with ${\rm U}(1)_Y$ and ${\rm U}(1)_{Y^{\bot}}$ are given by linear combinations of them. 
The assignment of the two Higgs doublets is determined so as 
to be compatible with that of the quarks originating from $\Psi_{+}(\repr{84})$; 
that is, the identification follows from the effective Yukawa  interactions 
of the quarks with the respective Higgs doublets, 
which are inherited from the higher-dimensional gauge interaction. 
If instead we use quarks from $\Psi_{-}(\repr{84})$, the roles of  $H_u^*$ and $H_d$ are 
interchanged. 
For leptons from $\Psi_{+}(\repr{36})$ and $\Psi_{-}(\repr{36})$, $E_{\rm R}$ couples to
$H_d$ and $H_u^*$, respectively.
In this way, the 4D effective theory of our model becomes 2HDM similarly to the SU(6) model,
and we point out that 
a model constructed from $\Psi_{+}(\repr{84})$ and $\Psi_{+}(\repr{36})$ is regarded as type II 2HDM 
and that constructed from $\Psi_{+}(\repr{84})$ and $\Psi_{-}(\repr{36})$ is considered as type Y (flipped) 2HDM \cite{BH&P,Grossman,AKT&Y,Branco:2011iw}.

For either type of 2HDMs, the EWSB vacuum is
analyzed in later section within the same framework as in the SU(6)
model discussed in subsection~\ref{sec:su6vac}; we restrict the
explicit analysis to the tree-level flat direction parametrized by the
Wilson line phases, and determine the vacuum from their
one-loop effective potential. In a more general low-energy
effective-theory treatment, loop-induced quadratic terms may also
induce small VEVs along non-flat directions, but such an analysis is
beyond the scope of the present work. We also assume that the effects
of localized tadpoles are sufficiently small, as discussed in
subsection~\ref{sec:su6tad}.

In this way, we find that leptons, quarks, the SM gauge bosons (and an
extra U(1) gauge boson) and two weak Higgs doublets originate from the
zero modes of three sets of 6D Weyl fermion in the $\bm{36}$
representation, three sets of 6D Weyl fermion in the $\bm{84}$
representation, the 4D components and the extra-dimensional components
of the 6D gauge boson in the $\bm{80}$ representation, respectively,
without generating an exotic type of quarks, leptons and scalar
fields. We note that 6D chiral anomalies appear for the above set of
fermions and a consistency of our model can be threatened. We provide
an anomaly free model in the next section.

%%%%%%%%%%%%%%%%%%%%%%%%%%%%%%%%%%%%%%%%%%%%%%%%%%%%%%%%%%%%%%%%
%
\section{Features of SU(9) grand unified model}
\label{sec:Features}
%
%%%%%%%%%%%%%%%%%%%%%%%%%%%%%%%%%%%%%%%%%%%%%%%%%%%%%%%%%%%%%%%%

%%%%%%%%%%%%%%%%%%%%%%%%%%%%%%%%%%%%%%%%%%%%%%%%%%%%
\subsection{Fermion content free of the bulk gauge anomalies}
\label{sec:Anomaly}
%%%%%%%%%%%%%%%%%%%%%%%%%%%%%%%%%%%%%%%%%%%%%%%%%%%%

In this subsection, we examine the gauge anomalies of the 6D bulk theory induced by the bulk fermion sector. 
Although reducible bulk anomalies could in principle be canceled by a Green-Schwarz (GS) mechanism, we adopt a more restrictive 
bottom-up viewpoint and require the bulk fermion content itself to be free of the bulk gauge anomalies. 
A complete fixed-point analysis of localized anomalies lies beyond the scope of the present work; 
when relevant, we discuss only the corresponding constraints in the 4D effective theory.

We consider a simple group G as a gauge group whose generators are $T^a$.
The 6D gauge anomaly is proportional to a group-theoretical factor such as
\beqn
\sum_{\Psi_\chi({\bf R})}\chi{\rm Str}_{\bf R}(T^{a_1}T^{a_2}T^{a_3}T^{a_4}),
\label{Eq:6Dgaugeanomaly}
\eeqn where $\Psi_\chi({\bf R})$ denotes a 6D Weyl fermion in
${\bf R}$ representation with 6D chirality $\chi (=\pm)$, and
${\rm Str}_{\bf R}$ stands for the trace over the symmetrized product
of gauge group generators of ${\bf R}$ representation.  From
eq.~\eqref{Eq:6Dgaugeanomaly}, we find that this trace is invariant
under the exchange between $T^a$ and $-(T^a)^*$, and hence the chiral
anomaly from a fermion in $\repr{\overline{R}}$ agrees with that from
a fermion in ${\bf R}$.  Furthermore, 6D fermions with a different
chirality have the anomaly coefficients with the opposite sign, and
then we can have an anomaly free set of 6D vector-like fermions.  In
this case, however, even numbers of families can appear from the set,
or exotic matter and/or mirror particles can come from it, depending
on the choice of $\eta_R$ and $\eta_T$. Then, extra brane fermions are
needed for three families to survive at the electroweak scale.  Therefore,
we search for an anomaly free set composed of 6D chiral fermions in
the following.

The 6D gauge anomaly relating to a 6D Weyl fermion in ${\bf R}$ follows the formula:
\beqn
{\rm Tr}_{\bf R}F^4 = \alpha_{\bf R} {\rm Tr}F^4 + c_{\bf R}({\rm Tr}F^2)^2,
\label{Eq:6Danomaly}
\eeqn
where ${\rm Tr}$ on the right-hand side are the trace of the fundamental representation, $F$ is the gauge field strength, and $\alpha_{\bf R}$ and $c_{\bf R}$ are the anomaly coefficients listed in table~\ref{Table:6Danomaly}~\cite{Erler:1993zy,Bhardwaj:2015xxa}.
%%%%%%%%%%%%%%%%%
\begin{table}[]
 \begin{center}
  \caption{6D anomaly coefficients for ${\bf R}$ representation of SU($n$) ($n \ge 4$)}
  \label{Table:6Danomaly}
  \begin{tabular}{c|ccccc}\hline
    ${\bf R}$ & $d_{\bf R}$ & $\alpha_{\bf R}$ & $c_{\bf R}$ & $l_{\bf R}$ & $A_{\bf R}$\\ \hline\hline
  fund  & $n$  & $1$ & $0$ & $1$ & $1$ \\ \hline
  asym  & $\frac{n(n-1)}{2}$ & $n-8$ & $3$ & $n-2$ & $n-4$ \\ \hline
  sym  & $\frac{n(n+1)}{2}$ & $n+8$ & $3$ & $n+2$ & $n+4$ \\ \hline
  adj  & $n^2 - 1$ & $2n$ & $6$ & $2n$ & $0$ \\ \hline
  rank-3  & $\frac{n(n-1)(n-2)}{6}$ & $\frac{n^2}{2} - \frac{17}{2}n + 27$ 
       & $3n-12$ & $\frac{n^2}{2} - \frac{5}{2}n + 3$ & $\frac{(n-3)(n-6)}2$ \\ \hline
  \end{tabular}
 \end{center}
\end{table}
%%%%%%%%%%%%%%%%%
In this table, the 6D anomaly coefficients are presented for a 6D fermion 
in ${\bf R}$ representation of SU($n$) ($n \ge 4$), and fund, asym, sym, adj and rank-3 stand for fundamental, anti-symmetric rank-2, symmetric rank-2, adjoint and anti-symmetric rank-3 tensor representations, respectively,
$d_{\bf R}$ is the dimension of representation and $l_{\bf R}$ is the index
defined as $\displaystyle{{\rm Tr}_{\bf R}(T^a)^2 = \frac{l_{\bf R}}{2}}$. 
Here, the 4D anomaly coefficient $A_{\bf R}$ defined by 
$\displaystyle{{\rm Tr}_{\bf R}(T^a)^3 = A_{\bf R}\displaystyle{{\rm Tr}(T^a)^3}}$ 
is also shown for the later convenience.
Specifically, the anomaly coefficients for fundamental,
anti-symmetric rank-2 and anti-symmetric rank-3 tensor representations of SU(9) are given
in table~\ref{Table:6DanomalySU9}.
%%%%%%%%%%%%%%%%%
\begin{table}[]
 \begin{center}
  \caption{6D anomaly coefficients for ${\bf R}$ representation of SU(9)}
  \label{Table:6DanomalySU9}
  \begin{tabular}{c|ccccc}\hline
    ${\bf R}$ & $d_{\bf R}$ & $\alpha_{\bf R}$ & $c_{\bf R}$ & $l_{\bf R}$ & $A_{\bf R}$\\ \hline\hline
  fund ($\repr{9}$) & $9$  & $1$ & $0$ & $1$ & $1$ \\ \hline
  asym ($\repr{36}$) & $36$ & $1$ & $3$ & $7$ & $5$ \\ \hline
  rank-3 ($\repr{84}$) & $84$ & $-9$ & $15$ & $21$ & $9$ \\ \hline
  \end{tabular}
 \end{center}
\end{table}
%%%%%%%%%%%%%%%%%

Using table~\ref{Table:6DanomalySU9}, we find that the 6D Weyl fermion content consisting of three copies
of $\repr{36} + \repr{84}$ with positive 6D chirality,
21 copies of $\repr{9} + \repr{\overline{9}}$ with positive chirality and 
9 copies of $\repr{36} + \repr{\overline{36}}$ with negative chirality is anomaly free,\footnote{
Other anomaly-free chirality assignments are possible. 
For instance, one of the three SM generations may be given the opposite 6D chirality; 
then the number of additional fields required for anomaly cancellation is reduced by a factor of three, 
although the flavor $\rm{SU}(3)$ symmetry is explicitly broken. 
Alternatively, one may flip the chirality of the entire $\repr{36}$ sector, 
thereby reducing the required number of $\repr{36}+\repr{\overline{36}}$ pairs from 9 to 6.} 
namely, the total contributions to $\alpha_{\bf R}$ and $c_{\bf R}$ vanish.
Furthermore, the mixed gauge-gravitational anomaly is also canceled, since 
the total contribution to $l_{\bf R}$ vanishes as a consequence of 
the relation $l_{\bf R} = \alpha_{\bf R} + 2c_{\bf R}$ for ${\bf R}=\repr{9},\repr{36},\repr{84}$. 
The pure gravitational anomaly, on the other hand, can be canceled out by adding suitable 6D chiral fermions in the SU(9) singlet representation.

Because the extra matters, namely 21 copies of $\repr{9} + \repr{\overline{9}}$ and 
9 copies of $\repr{36} + \repr{\overline{36}}$, are vector-like from the 4D point of view, 
they can acquire heavy masses and would not appear around the TeV scale.

%%%%%%%%%%%%%%%%%%%%%%%%%%%%%%%%%%%%%%%%%%%%%%%%%%%%
\subsection{Gauge coupling unification}
\label{sec:GCU}
%%%%%%%%%%%%%%%%%%%%%%%%%%%%%%%%%%%%%%%%%%%%%%%%%%%%

Let us examine whether the gauge coupling constants are unified or
not, under the assumption that the breaking of
${\rm U}(1)_{y}\times {\rm U}(1)_{\rm IV} \to {\rm U}(1)_Y$ due to a
VEV of a scalar field occurs at the same time as the orbifold breaking
${\rm SU}(9) \to {\rm SU}(2)_{\rm D}\times {\rm SU}(3)_{\rm C}\times
{\rm U}(1)_{y}\times {\rm U}(1)_{\rm IV}$
(${\rm SU}(2)_{\rm D}={\rm SU}(2)_{\rm L}$).  As an approximation, we
take an extra Higgs doublet to survive below the TeV scale.

We denote the 4D gauge coupling constant of the bulk SU(9) as $g_9$
and those of the SM gauge group SU(3)$_{\rm C}$, SU(2)$_{\rm L}$ and
U(1)$_Y$ as $g_{\rm s}$, $g$ and $g'$, respectively. Then, we find the
unification relation:
\beqn
g_9(M_{\rm U}) = g_{\rm s}(M_{\rm U}) =
\sqrt{2} g(M_{\rm U}) = \sqrt{\frac{14}{3}} g'(M_{\rm U}),
\label{Eq:unification}
\eeqn
where $M_{\rm U}$ is a unification scale and it is, in most cases, identified as the compactification scale $1/R$. In eq.~\eqref{Eq:unification}, the relation $g_9 = \sqrt{2} g$ is derived from the matching condition at $M_{\rm U}$:
\beqn
\frac{1}{g^2(M_{\rm U})} = \frac{1}{g_{\rm I}^2(M_{\rm U})} 
+ \frac{1}{g_{\rm II}^2(M_{\rm U})},~~
g_{9}(M_{\rm U}) = g_{\rm I}(M_{\rm U}) = g_{\rm II}(M_{\rm U}),
\label{Eq:g}
\eeqn
where $g_{\rm I}$ and $g_{\rm II}$ are the gauge coupling constants of SU(2)$_{\rm I}$ 
and SU(2)$_{\rm II}$, respectively.
In a similar way, the relation $g_9 = \sqrt{\frac{14}{3}} g'$ is derived from the matching condition $g' Q_Y = g_9 T_Y$ at $M_{\rm U}$ and $T_Y = \sqrt{\frac{3}{14}} Q_Y$.
From eq.~\eqref{Eq:unification}, we estimate the magnitude of the Weinberg angle $\theta_{\rm W}$ at $M_{\rm U}$ as
\beqn
\sin^2\theta_{\rm W}(M_{\rm U}) \equiv 
\frac{g'^2(M_{\rm U})}{g^2(M_{\rm U}) + g'^2(M_{\rm U})}
= \frac{3}{10}.
\label{Eq:thetaW}
\eeqn
This value is reasonably close to the observed value~\cite{ParticleDataGroup:2024cfk},
$\sin^2\theta_{\rm W}(M_Z) = 0.23122$, 
whereas the naive prediction $\sqrt3/2\simeq0.87$ of the SU(6) model is far from it. 
In fact, it is closer to the observed value than the ordinary GUT prediction, $3/8$,  
and hence it is expected that the unification scale is much lower than that of ordinary GUTs such as $10^{14}$--$10^{16}$~GeV. 

Here, we define modified fine-structure constants $\tilde\alpha_i$ ($i=2,1$) constructed from
$g_2 = \sqrt{2}g$ and $g_1 = \sqrt{14/3}g'$ as
\beqn
\tilde\alpha_2 \equiv \frac{g_2^2}{4\pi} = \frac{g^2}{2\pi},~~
\tilde\alpha_1 \equiv \frac{g_1^2}{4\pi} = \frac{7g'^2}{6\pi},
\eeqn
in addition to
$\displaystyle{{\tilde \alpha}_3 \equiv \alpha_3=
    \frac{g_{\rm s}^2}{4\pi}}$. Then,
the above unification relation is rewritten as
\beqn
\alpha_{\rm U}(M_{\rm U}) = \tilde\alpha_3(M_{\rm U}) = \tilde\alpha_2(M_{\rm U}) = \tilde\alpha_1(M_{\rm U}),
\label{Eq:unification-alpha}
\eeqn
where
$\displaystyle{\alpha_{\rm U} \equiv \frac{g_9^2}{4\pi}}$.

The renormalization group equations at the one-loop level are expressed by 
\beqn
\mu \frac{d}{d\mu} \tilde\alpha_i^{-1}(\mu) = - \frac{\tilde b_i}{2\pi},
\label{Eq:RGeq}
\eeqn
where $\mu$ is some energy scale and $\tilde b_i$ are the coefficients of beta functions.
The solutions of eq.~\eqref{Eq:RGeq} are given as
\beqn
\tilde\alpha_i^{-1}(\mu) = \tilde\alpha_i^{-1}(\mu_0) - \frac{\tilde b_i}{2\pi}\ln\frac{\mu}{\mu_0}.
\label{Eq:RGeq-sol}
\eeqn

In our model, $\tilde b_i$ are calculated as
\beqn
\tilde b_3 = -7,~~ \tilde b_2 = -\frac{3}{2},~~ \tilde b_1 = \frac{3}{2}.
\label{Eq:bi}
\eeqn
Using the experimental data in ref.~\cite{ParticleDataGroup:2024cfk}:
\beqn
\alpha_3(M_Z) = 0.1180,~~
\alpha^{-1}(M_Z) = 127.930,~~ \sin^2\theta_{\rm W}(M_Z) =0.23122,
\label{Eq:alphaMZ}
\eeqn
we obtain the following values,
\beqn
\tilde\alpha_3^{-1}(M_Z) &=& 8.475,~~
\tilde\alpha_2^{-1}(M_Z) = \frac{1}{2} \alpha^{-1}(M_Z) \sin^2\theta_{\rm W}(M_Z) = 14.79,~
\nonumber \\
\tilde\alpha_1^{-1}(M_Z) &=& \frac{3}{14} \alpha^{-1}(M_Z) \cos^2\theta_{\rm W}(M_Z) = 21.08,
\label{Eq:alphaiMZ}
\eeqn
where we use the relation $e=g\sin\theta_{\rm W}= g'\cos\theta_{\rm W}$.

Using eq.~\eqref{Eq:RGeq-sol}, we obtain the formula of unification scale:
\beqn
M_{\rm U} = M_Z \exp\Ls2\pi\frac{\tilde\alpha_i^{-1}(M_Z)-\tilde\alpha_j^{-1}(M_Z)}{\tilde b_i-\tilde b_j}\Rs,
\label{Eq:unificationscale}
\eeqn
by taking $\mu = M_Z =91.1880~{\rm GeV}$ and $\mu_0 = M_{\rm U}$.
When we take $i=3$ and $j=2$, the unification scale is estimated as $M_{\rm U} = 1.24 \times 10^5$~GeV and the fine-structure constant takes $\alpha_{\rm U}(M_{\rm U}) = {1}/{16.5}$.
In the same way, they are estimated as $M_{\rm U} = 1.01 \times 10^6$~GeV and $\alpha_{\rm U}(M_{\rm U}) = {1}/{18.9}$ for $i=3$ and $j=1$ and $M_{\rm U} = 4.75 \times 10^7$~GeV and $\alpha_{\rm U}(M_{\rm U}) = {1}/{17.9}$ for $i=2$ and $j=1$.

%%%%%%%%%%%%%%%%%%%%%
\begin{figure}[]
\centering
  \includegraphics[width=11cm,bb=0 0 380 250]{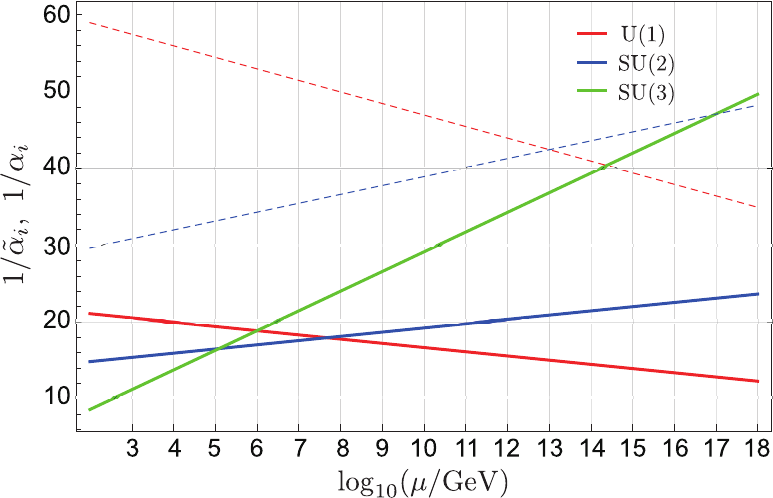} 
  \caption{Running of
    the modified fine-structure constants
    $\tilde \alpha_i$ $(i=1,2,3)$ in our model (solid lines).
    For comparison, the running of the fine structure constants
    $\alpha_i$ $(i=1,2,3)$ of the SM
    are also shown (dashed lines).
    The lines of $\tilde \alpha_3$ and $\alpha_3$ overlap.
    }\bigskip
\label{fig:GCU}
\end{figure}
%%%%%%%%%%%%%%%%%%%%%

Figure~\ref{fig:GCU} shows the running of gauge coupling constants in
our model (solid lines).  Here, we set the mass scale of the
additional Higgs doublet as the weak scale, for simplicity, since its
effect on the running is small enough.  Fine structure constants of
the SM, defined by $\alpha_3 = g_{\rm s}^2/(4\pi)$,
$\alpha_2 = g^2/(4\pi)$ and $\alpha_1 = 5/3 \cdot g'^2/(4\pi)$, are
also shown (dashed lines) for comparison. From figure~\ref{fig:GCU},
we can see that the gauge coupling constants take nearly the same
value around the scale of $10^6$~GeV in our model. Therefore, it is
expected that gauge coupling unification will be realized once various
effects, which have been neglected here for simplicity, are taken into
account. These can be classified into three types. First one is
threshold corrections due to Kaluza--Klein modes and vector-like
particles appearing in the cancellation of quantum anomalies. Their
magnitude can be estimated as
$O(\frac{\alpha_{\rm U}}{4\pi}\sum \ln(\Delta m_{\rm KK}/M_{\rm c}))$
and $O(\frac{\alpha_{\rm U}}{4\pi}\sum_k \ln(m_k/M_{Z}))$,
respectively, where $\Delta m_{\rm KK}$ are mass differences among
Kaluza--Klein modes, $M_{\rm c}(=O(1/R))$ is the compactification
scale, and $m_k$ are masses of vector-like particles.  Second one is
contributions from gauge kinetic terms localized on fixed points.
These terms can be induced by quantum corrections, and in principle a
value of $O(1)$ is allowed; however, in that case, the unification of
the gauge coupling constants could no longer be discussed due to lack
of predictability. Here, assuming that the gauge coupling constants
unify around the scale of $10^6$~GeV, we take this contribution to be
smaller than $O(1)$. Last one is higher-order (two-loop and beyond)
renormalization effects whose magnitudes can be negligibly small of
$O((\frac{\alpha_{\rm U}}{4\pi})^2)$. All of them depend on the
detailed structure of the model and lie beyond the scope of this
paper; they are left as subjects for future investigation.

%%%%%%%%%%%%%%%%%%%%%%%%%%%%%%%%%%%%%%%%%%%%%%%%%%%%
\subsection{Proton stability}
\label{sec:Proton}
%%%%%%%%%%%%%%%%%%%%%%%%%%%%%%%%%%%%%%%%%%%%%%%%%%%% 

Because the unification and compactification scales in the present
construction are relatively low, around $10^{6}$~GeV, it is therefore
crucial to examine whether the proton is stable enough or not to
explain the experimental bound $\tau_p > 10^{34}$~years
\cite{Super-Kamiokande:2020wjk}.

First of all, we point out that there are no proton decay processes
mediated by the gauge bosons, because the quarks belong to different
multiplets from the leptons in our model.

Next, we consider higher-dimensional operators which can induce proton decays, without specifying those origin. 
For such an operator with a mass dimension $d$ constructed from leptons and quarks, the proton life-time is roughly estimated as
\beqn
\tau_{p}^{(d)} \sim \frac{1}{\Gamma_{p}^{(d)}} \sim \left(\frac{\Lambda}{m_{p}}\right)^{2(d-4)}\frac{1}{m_{p}}
\sim \left(\frac{\Lambda}{m_{p}}\right)^{2(d-4)} \times 2.1 \times 10^{-32}~{\rm years},
\label{Eq:tau(d)}
\eeqn
where $\Gamma_{p}^{(d)}$ is the decay width of proton and $m_{p}$ is the proton mass.
Using eq.~\eqref{Eq:tau(d)}, $m_{p} \simeq 1$~GeV and $\Lambda \simeq 10^6$~GeV, the proton life-time becomes
\beqn
\tau_{p}^{(d)} \sim  2.1 \times 10^{12d-84}~{\rm years},
\label{Eq:tau(d)value}
\eeqn
and the mass dimension $d \ge 10$ is required to account for the experimental bound of the proton life-time $\tau_{p} \gtrsim 10^{34}$~years.

We can construct an operator with the mass dimension 6 from the bulk fields as
\beqn
O_{6} = U_{\rm R}U_{\rm R}D_{\rm R} E_{\rm R} \phi_{N}^{\dagger},
\label{operator-d=6}
\eeqn
where $\phi_{N}^{\dagger}$ is the hermitian conjugation of a scalar field having the same gauge quantum numbers as neutrino type lepton singlet $N_{\rm R}$.
This operator can induce a rapid proton decay.
In order to prevent such decay processes from occurring, it is necessary to forbid this operator. 

We note that, in contrast to conventional ${\rm SU(5)}$ unification, the present model allows us to impose a baryon-number symmetry 
${\rm U(1)_B}$ at the Lagrangian level. 
This is possible because quarks and leptons originate from different bulk fields, transforming in the $\bm{84}$ and $\bm{36}$, respectively, so that one may assign the ordinary baryon number $1/3$ to the quarks and $0$ to the leptons, thereby forbidding the dangerous proton-decay operators.
To protect baryon number against the effects of quantum gravity~\cite{Banks:2010zn, Harlow:2018tng}, 
we assume that ${\rm U(1)_B}$ is gauged, in which case the associated anomalies must be canceled. 

Concerning the 6D bulk anomalies, the mixed anomalies involving ${\rm SU(9)}$ and ${\rm U(1)_B}$ can be canceled 
without introducing a completely separate exotic sector. 
In particular, the additional fermions already required for the cancellation of the bulk ${\rm SU(9)}$ gauge anomaly can be assigned 
suitable ${\rm U(1)_B}$ charges so as to cancel these anomalies.
To be more concrete, when we denote the charges of the additional bulk fields transforming 
in the $\bm{9}$, $\overline{\bm9}$, $\bm{36}$ and $\overline{\bm{36}}$ representations as 
$q^i_{\bm{9}}$, $q^i_{\overline{\bm9}}$, $q^j_{\bm{36}}$ and $q^j_{\overline{\bm{36}}}$ ($i=1, \cdots, 21$, $j=1, \cdots, 9$), respectively, 
the conditions for the cancellation of the 
${\rm SU(9)}^3 {\rm U(1)}_{\rm B}$ and ${\rm SU(9)}^2 {\rm U(1)}_{\rm B}^2$ 
mixed anomalies 
are given by
\beqn
  \sum_{\Psi^\chi_{\bf R}}\chi A_{\bf R}q_{\Psi^\chi_{\bf R}} 
  &=& 9 + \sum_{i=1}^{21} (q^i_{\bm{9}} - q^i_{\overline{\bm9}}) 
   - 5 \sum_{j=1}^{9} (q^j_{\bm{36}} - q^j_{\overline{\bm{36}}}) = 0,
\label{Eq:9991}\\
  \sum_{\Psi^\chi_{\bf R}}\chi l_{\bf R}q_{\Psi^\chi_{\bf R}}^2 
  &=& 7 + \sum_{i=1}^{21} ((q^i_{\bm{9}})^2 + (q^i_{\overline{\bm9}})^2)
   - 7\sum_{j=1}^{9}((q^j_{\bm{36}})^2 + (q^j_{\overline{\bm{36}}})^2) = 0,
\label{Eq:9911}
\eeqn 
where $\chi=\pm$ denotes the 6D chirality and the group theoretical factors $A_{\bf R}$ and $l_{\bf R}$ are given 
in table~\ref{Table:6DanomalySU9} with $A_{\overline{\bf R}}=-A_{\bf R}$ and $l_{\overline{\bf R}}=l_{\bf R}$. 
Since fermions in $\bm{36}$ and $\overline{\bm{36}}$ contain zero modes, as discussed in subsection~\ref{sec:ZeroModes}, 
whereas those in $\bm{9}$ and $\overline{\bm{9}}$ need not, the former should carry opposite ${\rm U(1)_B}$ charges, 
$q^j_{\overline{\bm{36}}}=-q^j_{\bm{36}}$, so as to allow heavy mass terms. 
To demonstrate that the above equations admit a solution, we present the following explicit charge assignment: 
four of the charges $q^i_{\bm{9}}$ and three of the charges $q^i_{\overline{\bm{9}}}$ are set to $+1$; 
one of the charges $q^j_{\bm{36}}$ is set to $+1$, so that the corresponding charge is fixed to 
$q^j_{\overline{\bm{36}}}=-1$; all the remaining charges vanish.
The 6D gauge anomaly of ${\rm SU(9)} {\rm U(1)}_{\rm B}^3$ is absent from the traceless condition relating to the generators of SU(9).
The 6D gauge anomaly of ${\rm U(1)}_{\rm B}^4$ and ${\rm U(1)}_{\rm B}$-gravitational mixed anomalies can be canceled 
by adding SU(9) singlets with suitable baryon numbers, which may have zero modes or not.

As mentioned above, we do not attempt a complete fixed-point analysis of localized anomalies on $T^2/\Z4$ in the present work. 
We therefore work in the 4D effective-theory and focus on the globally nonvanishing part of the anomaly that survives 
after compactification.\footnote{Following the general analysis of localized anomalies on orbifolds, we assume that the globally vanishing part is 
canceled by a bulk four-form GS mechanism, whereas the globally nonvanishing part is canceled by a bulk two-form GS 
mechanism~\cite{Asaka:2002my, vonGersdorff:2003dt, vonGersdorff:2006nt}.
}
The remaining mixed anomaly in the 4D effective theory is assumed to be canceled by a GS mechanism, 
so that the anomalous ${\rm U(1)_B}$ gauge boson acquires a St\"uckelberg mass~\cite{Ibanez:1998qp,Anastasopoulos:2006cz}.
As a consequence, ${\rm U(1)_B}$ does not remain as an exact massless gauge symmetry, but the dangerous proton-decay operators 
are still perturbatively forbidden by the underlying gauge invariance. 
Baryon-number violation can then arise through nonperturbative effects and is therefore expected to be exponentially suppressed 
at low energies~\cite{tHooft:1976rip,Rubakov:1996vz}.\footnote{If one wishes to impose an exact low-energy selection rule forbidding the dangerous low-dimensional operators, 
i.e.\ those with $\Delta B<3$, one may further assume that an exact discrete remnant, such as baryon triality, 
survives below the St\"uckelberg scale~\cite{Ibanez:1991hv,Dreiner:2005rd}.}

In this way, our grand unified model has no perturbative operators
that induce proton decay and predicts null results of the proton decay
search, unless the breaking of ${\rm U(1)}_{\rm B}$ gauge symmetry is
unexpectedly large.

%%%%%%%%%%%%%%%%%%%%%%%%%%%%%%%%%%%%%%%%%%%%%%%%%%%%
\subsection{Electroweak symmetry breaking}
\label{sec:effpotsu9}
%%%%%%%%%%%%%%%%%%%%%%%%%%%%%%%%%%%%%%%%%%%%%%%%%%%%
As discussed in section~\ref{sec:SU9GUT}, the SU(9) model contains two
Higgs doublets as in the SU(6) case. These Higgs doublets come from
the zero modes of $A_z$ and $A_{\bar z}$ and contain the Wilson line
phase d.o.f. parametrized by two real parameters $a$ and $b$.  In the
field space of the Wilson line phases, there is a specific point
$(a,b)=(0,1/2)$, where the electroweak symmetry is realized.
Away from this point, EWSB occurs via the Hosotani mechanism.
We therefore discuss one-loop effective potentials for the
phases and EWSB.

From eqs.~\eqref{eq:SU(9)-w/WL} and~\eqref{Eq:<Az>}, we see that 
the Wilson line phases lie entirely in the SU(6) sector and are neutral under the color SU(3) subgroup.
Thus, the contributions to the
one-loop effective potential from bulk SU(9) multiplets are written
by the contributions in the SU(6) model summarized in
subsection~\ref{sec:su6vac}. In the following, we discard
  irrelevant constants that are independent of the Wilson line phases
  in the contributions to the effective potential. 

We first consider the contribution from a bulk field
in the fundamental representation $\repr{9}$. From the
decomposition in eq.~\eqref{Eq:decom9(6+3)-SU3}, under the SU(6)
subgroup of SU(9), $\repr{9}$ transforms as one $\repr{6}$ and three
singlets.  Thus, the effective potential contribution from a bulk
field in $\repr{9}$ is the same as the contribution from a bulk field
in $\repr{6}$ in the SU(6) model since three singlets in
  $\repr{9}$ do not couple to the Wilson line phases.  Let
$\hat {\cal V}_{\bf R}^{[\beta_T]}(a,b)$ be an effective potential
contribution from a real bosonic d.o.f. of a bulk field belonging to
an irreducible representation $\bf{R}$ in the SU(9) model. As in
the SU(6) case,
$\hat {\cal V}_{\repr{\overline{R}}}^{[\beta_T]}(a,b)=\hat {\cal
  V}_{\repr{R}}^{[\beta_T]}(a,b)$ holds, and the 
contributions are independent of a choice of $\eta_R$, while a
  choice of $\eta_T$ changes the contributions through a value of
  $\beta_T$.
We find
\begin{align}
  \hat {\cal V}^{[\beta_T]}_{\repr9}(a,b)
  &=  {\cal V}^{[\beta_T]}_\repr6(a,b), 
\end{align}
where ${\cal V}^{[\beta_T]}_\repr6(a,b)$ is shown in
eq.~\eqref{Eq:V-6}.

%%%%%%%%%%%%%%%%%%%%%
\begin{figure}[]
\centering
  \includegraphics[width=5cm,clip]{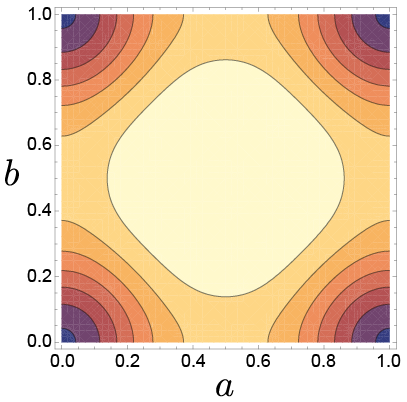} 
  \caption{The contour plot of the one-loop effective potential in the
    pure Yang-Mills case as a function of the Wilson line phases $a$
    and $b$ is depicted. The horizontal (vertical) axis shows a value of
    $a$ ($b$). From the light orange region to the dark blue region,
    values of the potential decrease.  }\bigskip
\label{fig:pYMpot}
\end{figure}
%%%%%%%%%%%%%%%%%%%%%

Using similar discussions, effective potential contributions from bulk
fields in other representations are easily derived. From the
decompositions in eqs.~\eqref{Eq:decom84-SU3}, \eqref{Eq:decom36-SU3}
and~\eqref{Eq:decom80-SU3}, we obtain
\begin{align}
  \hat {\cal V}^{[\beta_T]}_{\repr{36}}(a,b)
  &=  {\cal V}^{[\beta_T]}_{\repr{15}}(a,b)+3\times {\cal V}^{[\beta_T]}_{\repr6}(a,b),\\
  % \hat {\cal V}^{[\beta_T]}_{\repr{45}_S}(a,b)
  % &=  {\cal V}^{[\beta_T]}_{\repr{21}}(a,b)+3\times {\cal V}^{[\beta_T]}_{\repr6}(a,b),\\
  \hat {\cal V}^{[\beta_T]}_{\repr{80}}(a,b)
  &=  {\cal V}^{[\beta_T]}_{\repr{35}}(a,b)+6\times {\cal V}^{[\beta_T]}_{\repr6}(a,b),\\
  \hat {\cal V}^{[\beta_T]}_{\repr{84}}(a,b)
  &=  {\cal V}^{[\beta_T]}_{\repr{20}}(a,b)+3\times {\cal V}^{[\beta_T]}_{\repr{15}}(a,b)
    +3\times {\cal V}^{[\beta_T]}_{\repr{6}}(a,b), 
\end{align}
where the terms in the right-hand sides are shown in eqs.~\eqref{Eq:V-6}--\eqref{Eq:V-20}.
The numerical coefficients in front of the SU(6) contributions simply count the multiplicities of the corresponding SU(6) representations in the SU(9) decomposition.

  In figure~\ref{fig:pYMpot}, we show the contour plot of the one-loop
effective potential in the pure Yang-Mills case as a function of the
Wilson line phases $a$ and $b$. The potential is invariant under
integer shifts, sign flips and the exchange of $a$ and $b$.
The figure shows that, without contributions from bulk matter fields, the global
minimum appears at $(a,b)=(0,0)$, where the SU(9) symmetry is broken
down to U(1)$^4 \times$SU(4).
Therefore, in order to realize a vacuum near the special point $(a,b)=(0,1/2)$, which is relevant for EWSB in the present model, bulk matter contributions are essential.

%%%%%%%%%%%%%%%%%%%%%
\begin{figure}[]
\centering
  \includegraphics[width=5cm,clip]{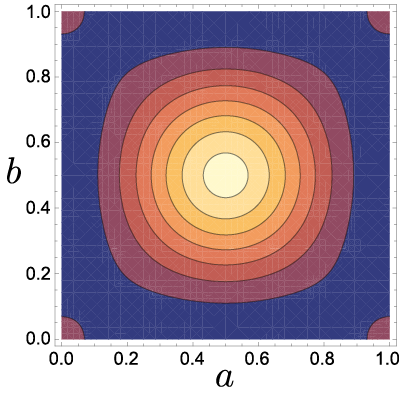}\qquad
  \includegraphics[width=5cm,clip]{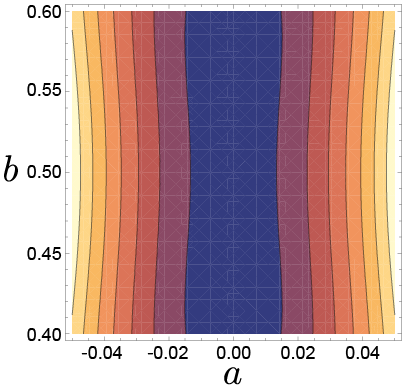} 
  \caption{The contour plot of the one-loop effective potential with a
    set of bulk matter fields as a function of the Wilson line phases
    $a$ and $b$ is depicted. The horizontal (vertical) axis shows a
    value of $a$ ($b$). From the light orange region to the dark blue
    region, values of the potential decrease.  In this case, the
    numbers of bulk 6D Weyl fermions are    $(n_{\repr{9}}^{[0]},n_{\repr{9}}^{[1/2]},n_{\repr{36}}^{[0]},n_{\repr{36}}^{[1/2]})
    =(21,21,3,3)$.  }\bigskip
\label{fig:wMpot}
\end{figure}
%%%%%%%%%%%%%%%%%%%%%

As an illustrative benchmark, we show in figure~\ref{fig:wMpot} the
one-loop effective potential including a suitable set of bulk matter
fields. Let $n_{\bf{R}}^{[\beta_T]}$ denote the number of 6D Weyl
fermions in the SU(9) representation $\repr{R}$ with translation phase
$\beta_T$. In the example shown in the figure, we take
$(n_{\repr{9}}^{[0]},n_{\repr{9}}^{[1/2]},n_{\repr{36}}^{[0]},n_{\repr{36}}^{[1/2]})
=(21,21,3,3)$.  These fermions may be regarded as part of an
anomaly-free bulk fermion set discussed in
subsection~\ref{sec:Anomaly}.  For the purpose of this illustrative
analysis, the additional fermions required for anomaly
cancellation are assumed to give negligible contributions to the
  effective potential for the Wilson line phases. This
  situation may be realized, for example, when these additional
  fermions become sufficiently heavy. For this matter
content, the potential develops degenerate minima near
$(a,b)\simeq (0,1/2\pm 0.076)$, and EWSB occurs. Although the
displacement from the special point is relatively small, it is still
not sufficient to generate a realistic hierarchy between the
electroweak scale and the compactification scale in this simple
benchmark.
By including appropriately chosen additional bulk matter fields, the displacement can be further reduced. For example, we
  have confirmed that 
  when several copies of a matter set characterized by
$(n_{\repr{9}}^{[0]},n_{\repr{9}}^{[1/2]},n_{\repr{36}}^{[0]},n_{\repr{36}}^{[1/2]})
=(5,5,1,1)$ 
are added to the benchmark matter content,
the potential develops vacua with
$(a,b)\simeq (0,1/2\pm {\cal O}(10^{-2}))$,
but the resulting displacement remains insufficient to reproduce the phenomenologically required hierarchy, ${\cal O}(10^{-3}$--$10^{-5})$.
A realistic hierarchy would therefore require further model building beyond this illustrative benchmark, possibly including an adjustment of the bulk matter content, some tuning of continuous parameters in a realistic model and a detailed treatment of threshold corrections.

%%%%%%%%%%%%%%%%%%%%%%%%%%%%%%%%%%%%%%%%%%%%%%%%%%%%%%%%%%%%%%%%
%
\section{Conclusions and discussions}
\label{sec:concl}
%
%%%%%%%%%%%%%%%%%%%%%%%%%%%%%%%%%%%%%%%%%%%%%%%%%%%%%%%%%%%%%%%%

We have studied a 6D SU(9) grand unified model on the
orbifold $T^2/\Z4$ with rank-reducing discrete BCs and
continuous Wilson line phases.  The present model is a grand unified
extension of the SU(6) gauge-Higgs unification model~\cite{Kawamura:2025}, in which the
electroweak symmetry structure, two Higgs doublets and quark zero modes
were realized by the same type of BCs and Wilson line
dynamics.

The essential idea of the present construction is to embed the SU(6)
electroweak gauge-Higgs unification sector into SU(9), which contains
${\rm SU}(6)\times {\rm SU}(3)\times {\rm U}(1)$ as a subgroup.  The
additional SU(3) factor is identified with the color gauge symmetry,
while the extra Abelian generator plays an important role in the
embedding of the weak hypercharge.  Around the special point
$(a,b)=(0,1/2)$ in the field space of the Wilson line
phases, the gauge symmetry is enhanced to
${\rm SU}(2)_{\rm D}\times {\rm SU}(3)_{\rm C}\times {\rm
  U}(1)_y\times {\rm U}(1)_{\rm IV}$.  The SM hypercharge is realized
as a linear combination of ${\rm U}(1)_y$ and ${\rm U}(1)_{\rm IV}$.
We assume that the ${\rm U}(1)_y\times {\rm U}(1)_{\rm IV}$ symmetry
is broken down to ${\rm U}(1)_Y$ around the compactification scale by
the VEV of a scalar field which is a singlet under the SM gauge group
${\rm SU}(2)_{\rm D}\times {\rm SU}(3)_{\rm C}\times {\rm U}(1)_Y$.

We have shown that the SM matter fields can be obtained from bulk
fields without exotic zero modes.  In particular, the quarks in each
generation arise from a bulk fermion in the $\bm{84}$ representation of
SU(9), while the leptons arise from a bulk fermion in the $\bm{36}$
representation.  This is a significant improvement over the SU(6) model,
where the quarks could be accommodated naturally but the leptons were not
included in the same minimal setup.  The two Higgs doublets originate from the
zero modes of the extra-dimensional components of the SU(9) gauge field, as
in the SU(6) model.
A model constructed from $\Psi_{+}(\repr{84})$ and $\Psi_{+}(\repr{36})$ is regarded as type II 2HDM and that constructed from $\Psi_{+}(\repr{84})$ and $\Psi_{-}(\repr{36})$ is considered as type Y (flipped) 2HDM.

We have also discussed several consistency and phenomenological
aspects of the model.  We constructed matter contents which are free
from 6D bulk SU(9) gauge anomalies.  The additional fields introduced
for anomaly cancellation can be vector-like from the 4D point of view.
We also examined the gauge coupling relations.  Owing to the modified
weak hypercharge embedding in the SU(9) model, the tree-level
prediction for the Weinberg angle is improved compared with the
original SU(6) model.  The magnitude of the Weinberg angle is
calculated as $\sin^2\theta_{\rm W}(M_{\rm U}) = 3/10$ at the
unification scale $M_{\rm U}$.  This value is smaller than the
conventional ${\rm SU}(5)$ GUT prediction such as
$\sin^2\theta_{\rm W}(M_{\rm U}) =3/8$, and hence it is anticipated
that $M_{\rm U}$ is much lower than that of ordinary GUTs such as
$10^{14}$--$10^{16}$~GeV.  In our model, $M_{\rm U}$ is estimated as
$10^{5}$--$10^{7}$~GeV.  Precise gauge coupling unification, however,
is expected to depend on threshold corrections, brane-localized gauge
kinetic terms and the detailed mass spectrum of additional fields.

A characteristic feature of the model is the absence of proton decay
mediated by bulk gauge bosons, which is seen from the fact that the
quarks and leptons are embedded into different bulk multiplets.  To
control higher-dimensional baryon-number-violating operators, we
introduce a gauged baryon-number symmetry, U(1)$_{\rm B}$, which is
consistent with our GUT symmetry, in contrast to ordinary SU(5)
unification.  While its 6D bulk anomalies can be cancelled, its 4D
anomaly implies that U(1)$_{\rm B}$ would not remain as an exact
massless gauge symmetry.  We assume that, because of the underlying
gauge invariance, baryon-number-violating terms can arise only through
nonperturbative effects and are therefore exponentially suppressed.
Despite the smaller unification scale, our model then predicts null
results in proton-decay searches, unless the breaking of
${\rm U(1)}_{\rm B}$ gauge symmetry is unexpectedly large.

We have further studied the EWSB through the Hosotani mechanism.  The
one-loop effective potential for the Wilson line phases can be
expressed in terms of contributions closely related to those in the
SU(6) model.  By adding suitable bulk matter fields, we found examples
in which the minimum of the effective potential is slightly displaced
from $(a,b)=(0,1/2)$, leading to the EWSB.  These examples show that
the desired electroweak vacuum can be realized in the present SU(9)
framework.

The examples found in this paper should, however, be regarded as toy
models.  In the numerical examples, the compactification scale
$M_{\rm c}(=1/R)$ is only about $10$--$100$ times larger than the
electroweak scale $M_{\rm EW}(=v)$.  To obtain a more realistic
hierarchy, such as ${M_{\rm c}}/{M_{\rm EW}} ={\cal O}( 10^3$--$10^5)$, a
further tuning of the bulk matter contents and parameters would be
necessary.  This situation is similar to the usual little hierarchy
problem in gauge-Higgs unification models.  A systematic search for
matter contents that realize a larger hierarchy is an important
subject for future work.

Several other issues remain open.  In this paper, we have revisited
the issue of tadpole terms for the field strength $F_{56}$ localized
at orbifold fixed points and found that the argument presented in
ref.~\cite{Kawamura:2025} is not sufficient to resolve this issue.  We
therefore assume that the tadpole contributions are sufficiently
small.  A natural mechanism for suppressing these contributions is
still needed.  Localized anomalies should also be analyzed
to establish a fully consistent ultraviolet description.  It
is also important to construct realistic fermion mass matrices,
including flavor mixing and neutrino masses, possibly by using
brane-localized interactions or mixing with vector-like fermions.  The
effects of brane-localized operators on gauge coupling unification,
flavor physics and the Higgs sector should also be studied.

The present work demonstrates that rank-reducing discrete BCs
on $T^2/\Z4$ provide a useful framework for extending electroweak
gauge-Higgs unification to a grand unified setting.  Compared with the
previous SU(6) model, the SU(9) model incorporates color interactions,
leptons, a viable weak hypercharge embedding, anomaly-free bulk matter contents 
and a structural suppression of gauge-boson-mediated proton decay.  Although
a realistic electroweak hierarchy still requires further model building, the
results obtained here suggest that the combination of rank-reducing BCs 
and Wilson line dynamics is a promising direction for constructing
higher-dimensional unified theories.

%%%%%%%%%%%%%%%%%%%%%%%%%%%%%%%%%%%%%%%%%%%%%%%%%%%%%%%%%%%%%%%% 
\section*{Acknowledgments}
The authors would like to thank Koji Tsumura for helpful comments.
This work was supported in part by scientific grants from the Ministry of Education, Culture, Sports, Science and Technology under Grant No.~22K03632 (YK).
%%%%%%%%%%%%%%%%%%%%%%%%%%%%%%%%%%%%%%%%%%%%%%%%%%%%%%%%%%%%%%%%

\appendix

%%%%%%%%%%%%%%%%%%%%%%%%%%%%%%%%%%%%%%%%%%%%%%%%%%%%%%%%%%%%%%%%
%
\section{Study based on ${\rm SU}(2)\sub{I}\times {\rm SU}(2)\sub{II}\times {\rm SU}(4)\times {\rm U}(1)_A
\times {\rm U}(1)_B\times {\rm U}(1)_C$}
\label{app:SU4}
%
%%%%%%%%%%%%%%%%%%%%%%%%%%%%%%%%%%%%%%%%%%%%%%%%%%%%%%%%%%%%%%%%

First, we present two kinds of decompositions of SU(9) such as
\beqn
{\rm SU}(9) &\supset& {\rm SU}(6) \times {\rm SU}(3) \times {\rm U}(1)_{\rm IV}
\nonumber\\
&\supset& {\rm SU}(2)\sub{I}\times {\rm SU}(2)\sub{II}\times {\rm U}(1)_y\times {\rm U}(1)_A\times {\rm U}(1)\sub{III}\times {\rm SU}(3)\times {\rm U}(1)\sub{IV},
\label{Eq:decomposition1}
\eeqn
and
\beqn
{\rm SU}(9) &\supset& 
{\rm SU}(2)_{\rm I}\times {\rm SU}(2)_{\rm II}\times {\rm SU}(4)\times {\rm U}(1)_A\times {\rm U}(1)_B\times {\rm U}(1)_C
\nonumber\\
&\supset& {\rm SU}(2)_{\rm I}\times {\rm SU}(2)_{\rm II}\times {\rm SU}(3)\times {\rm U}(1)_A\times {\rm U}(1)_B\times {\rm U}(1)_C\times {\rm U}(1)_D.
\label{Eq:decomposition2}
\eeqn
We note that the SU(9) symmetry is broken down to ${\rm SU}(2)_{\rm I}\times {\rm SU}(2)_{\rm II}\times {\rm SU}(4)\times {\rm U}(1)_A\times {\rm U}(1)_B\times {\rm U}(1)_C$ by $R_0$ given in \Eqref{Eq:SU(9)R0-permute}. The generators of ${\rm U}(1)_y\times {\rm U}(1)\sub{III}\times {\rm U}(1)\sub{IV}$ are realized by linear combinations of those of ${\rm U}(1)_B\times {\rm U}(1)_C\times {\rm U}(1)_D$, as given in \Eqref{Eq:SU(9)GHU-RelsU(1)}.
In this appendix, we examine the particle assignment based on the second decomposition.

Under ${\rm SU}(2)\sub{I}\times {\rm SU}(2)\sub{II}\times {\rm SU}(4)\times {\rm U}(1)_A\times {\rm U}(1)_B\times {\rm U}(1)_C$,
$\repr9$, $\repr{\overline{9}}$, $\repr{36}$ and $\repr{84}$ of SU(9) are decomposed into
\beqn
\repr9&=&(\repr2, \repr1, \repr1)^2_{1,1,4}+(\repr1, \repr2, \repr1)^{0}_{-1,1,4}
+ (\repr1, \repr1, \repr1)^{1}_{0,-4,4} + (\repr1, \repr1, \repr4)^{-1}_{0,0,-5},
\label{Eq:decom9}\\
\repr{\overline{9}}&=& (\repr2, \repr1, \repr1)^2_{-1,-1,-4} + (\repr1, \repr2, \repr1)^{0}_{1,-1,-4}
+ (\repr1, \repr1, \repr1)^{-1}_{0,4,-4} + (\repr1, \repr1, \repr{\overline{4}})^{1}_{0,0,5},
\label{Eq:decom9bar}\\
\repr{36}&=&(\repr1, \repr1, \repr1)^{0}_{2,2,8}+(\repr1, \repr1, \repr1)^{0}_{-2,2,8}+(\repr2, \repr2, \repr1)^2_{0,2,8}+(\repr2, \repr1, \repr1)^{-1}_{1,-3,8}+(\repr1, \repr2, \repr1)^{1}_{-1,-3,8}
\nonumber \\
&~& +(\repr2, \repr1, \repr4)^1_{1,1,-1}+(\repr1, \repr2, \repr4)^{-1}_{-1,1,-1}
+(\repr1, \repr1, \repr4)^{0}_{0,-4,-1}+(\repr1, \repr1, \repr6)^{2}_{0,0,-10},
\label{Eq:decom36}\\
\repr{84}&=&(\repr1, \repr1, \repr4)^{-1}_{2,2,3}+(\repr1, \repr1, \repr4)^{-1}_{-2,2,3}+(\repr2, \repr1, \repr4)^2_{1,-3,3}+(\repr1, \repr2, \repr4)^{0}_{-1,-3,3}
\nonumber \\
&~& +(\repr1, \repr2, \repr6)^{2}_{-1,1,-6}+(\repr2, \repr1, \repr6)^{0}_{1,1,-6}
+ (\repr1, \repr1, \repr6)^{-1}_{0,-4,-6}+(\repr1, \repr1, \repr{\overline{4}})^{1}_{0,0,-15}
\nonumber \\
&~& +(\repr1, \repr1, \repr1)^{1}_{2,-2,12}+(\repr1, \repr1, \repr1)^{1}_{-2,-2,12}
+(\repr2, \repr2, \repr1)^{-1}_{0,-2,12}+(\repr2, \repr2, \repr4)^{1}_{0,2,3}
\nonumber \\
&~& +(\repr2, \repr1, \repr1)^2_{-1,3,12}+(\repr1, \repr2, \repr1)^{0}_{1,3,12},
\label{Eq:decom84}
\eeqn
where representations/charges of $({\rm SU}(2)\sub{I}, {\rm SU}(2)\sub{II}, {\rm SU}(4))^{R_0}_{{\rm U}(1)_A, {\rm U}(1)_B, {\rm U}(1)_C}$
are shown at the indicated positions.

Using the observation that no zero modes appear from bulk fields which belong to odd-rank tensor multiplets of SU(6) and assigning suitable values to $\eta_R$ and $\eta_T$, we can obtain quarks and leptons as zero modes of $\repr{84}$ and $\repr{36}$, respectively.
We explain it by taking $\repr{36}$ as an example.
The linear combination 
$(\repr1, \repr1, \repr1)^{0}_{2,2,8}-(\repr1, \repr1, \repr1)^{0}_{-2,2,8}$
and the submultiplet $(\repr1, \repr1, \repr1)^{0}_{0,-4,-1}$ in $(\repr1, \repr1, \repr4)^{0}_{0,-4,-1}$ contain zero modes for $\eta_R = 1$ and $\eta_T =-1$.
The linear combination of
$(\repr1, \repr2, \repr1)^{1}_{-1,-3,8}$ and the submultiplet 
$(\repr2, \repr1, \repr1)^1_{1,1,-1}$ in $(\repr2, \repr1, \repr4)^1_{1,1,-1}$
contains a zero mode for $\eta_R =-i$  and $\eta_T =-1$.
For $\eta_R = 1$ or $\eta_R = -i$, no zero modes appear from $(\repr2, \repr2, \repr1)^2_{0,2,8}$,
$(\repr2, \repr1, \repr1)^{-1}_{1,-3,8}$,
$(\repr1, \repr2, \repr4)^{-1}_{-1,1,-1}$ and $(\repr1, \repr1, \repr6)^{2}_{0,0,-10}$.
Furthermore, no zero modes stem from the submultiplet $(\repr2, \repr1, \repr3)^1_{1,1,-1}$ in
$(\repr2, \repr1, \repr4)^1_{1,1,-1}$ and the submultiplet $(\repr1, \repr1, \repr3)^{0}_{0,-4,-1}$ in $(\repr1, \repr1, \repr4)^{0}_{0,-4,-1}$ in addition to the submultiplet $(\repr1, \repr2, \repr3)^{-1}_{-1,1,-1}$ in $(\repr1, \repr2, \repr4)^{-1}_{-1,1,-1}$, since they belong to an odd-rank tensor multiplet of SU(6).
The submultiplet $(\repr1, \repr1, \repr{\overline{3}})^{2}_{0,0,-10}$ in $(\repr1, \repr1, \repr6)^{2}_{0,0,-10}$ does not contain zero modes for $\eta_T = -1$.
For $\eta_R=i$, the linear combination 
$(\repr2, \repr1, \repr1)^{-1}_{1,-3,8}-(\repr1, \repr2, \repr1)^{-1}_{-1,1,-1}$ 
contains a zero mode where $(\repr1, \repr2, \repr1)^{-1}_{-1,1,-1}$ is a submultiplet of $(\repr1, \repr2, \repr4)^{-1}_{-1,1,-1}$.
Zero modes of $\repr{84}$ can also be explored in a similar manner.

Thus we find two types of particle assignments for quarks and leptons such that
\beqn
\Psi_{+}(\repr{84}) =\begin{cases}
 \psi_{(+,{\rm R})}(\repr{84}) \ni \begin{cases}
\MyDecomC{1}{1}{3}{2}{2}{3}{1}{-1} - \MyDecomC{1}{1}{3}{-2}{2}{3}{1}{-1} \Rightarrow D_{\rm R} \\
  \MyDecomC{1}{1}{3}{0}{-4}{-6}{-2}{-1} \Rightarrow U_{\rm R}
 \end{cases} \\
 \psi_{(+,{\rm L})}(\repr{84}) \ni \MyDecomC{2}{1}{3}{1}{1}{-6}{-2}{0} -\MyDecomC{1}{2}{3}{-1}{-3}{3}{1}{0}
\Rightarrow Q_{\rm L},
\end{cases}
\label{Eq:quarks+}
\eeqn
or
\beqn
\Psi_{-}(\repr{84}) =\begin{cases}
 \psi_{(-,{\rm R})}(\repr{84}) \ni \begin{cases}
  \MyDecomC{1}{1}{3}{2}{2}{3}{1}{-1} -\MyDecomC{1}{1}{3}{-2}{2}{3}{1}{-1} \Rightarrow D_{\rm R} \\
  \MyDecomC{1}{1}{3}{0}{-4}{-6}{-2}{-1} \Rightarrow U_{\rm R}
 \end{cases} \\
  \psi_{(-,{\rm L})}(\repr{84}) \ni
 \MyDecomC{2}{1}{3}{1}{-3}{3}{1}{2} - \MyDecomC{1}{2}{3}{-1}{1}{-6}{-2}{2}
\Rightarrow Q_{\rm L},
\end{cases}
\label{Eq:quarks-}
\eeqn
and
\beqn
\Psi_{+}(\repr{36}) =\begin{cases}
 \psi_{(+,{\rm R})}(\repr{36}) \ni \begin{cases}
 \MyDecomC{1}{1}{1}{2}{2}{8}{0}{0} - \MyDecomC{1}{1}{1}{-2}{2}{8}{0}{0} \Rightarrow E_{\rm R} \\
  \MyDecomC{1}{1}{1}{0}{-4}{-1}{-3}{0} \Rightarrow N_{\rm R}
 \end{cases} \\
 \psi_{(+,{\rm L})}(\repr{36}) \ni \MyDecomC{2}{1}{1}{1}{1}{-1}{-3}{1} - \MyDecomC{1}{2}{1}{-1}{-3}{8}{0}{1}
\Rightarrow L_{\rm L} ,
\end{cases}
\label{Eq:leptons+}
\eeqn
or
\beqn
\Psi_{-}(\repr{36}) =\begin{cases}
 \psi_{(-,{\rm R})}(\repr{36}) \ni \begin{cases}
  \MyDecomC{1}{1}{1}{2}{2}{8}{0}{0} - \MyDecomC{1}{1}{1}{-2}{2}{8}{0}{0} \Rightarrow E_{\rm R} \\
  \MyDecomC{1}{1}{1}{0}{-4}{-1}{-3}{0} \Rightarrow N_{\rm R}
 \end{cases} \\
  \psi_{(-,{\rm L})}(\repr{36}) \ni
  \MyDecomC{2}{1}{1}{1}{-3}{8}{0}{-1} - \MyDecomC{1}{2}{1}{-1}{1}{-1}{-3}{-1}
\Rightarrow L_{\rm L} ,
\end{cases}
\label{Eq:leptons-}
\eeqn
where representations/charges of 
$\Ls {\rm SU}(2)\sub{I},{\rm SU}(2)\sub{II}, {\rm SU}(3)\Rs_{{\rm U}(1)_A,{\rm U}(1)_B,{\rm U}(1)_C,{\rm U}(1)_D}^{R_0}$ are shown at the indicated positions, and ${\rm U}(1)_D$ is the U(1) subgroup of SU(4) whose generator is defined by \Eqref{Eq:SU(9)GHU-U(1)D}.
In identifying with the SM fermions, we use the definition of the weak hypercharge as
\beqn
Q_Y &\equiv& -\frac{1}{10}t_B -\frac{1}{10}t_C  + \frac{1}{6}t_D
  = -\frac{2}{\sqrt{10}}T_B -\frac{6}{\sqrt{10}}T_C +  \frac{2}{\sqrt{6}}T_D
\nonumber \\
  &=&{\rm diag}(-1/2,-1/2,-1/2,-1/2,0,0,2/3,2/3,2/3).
\label{Eq:hypercharge-SU9-BCD}
\eeqn

In a similar way, the adjoint representation \repr{80} of SU(9) is decomposed into
\beqn
 \repr{80}&=& (\repr3, \repr1, \repr1)^{0}_{0,0,0}+(\repr1, \repr3, \repr1)^{0}_{0,0,0}
 +(\repr1, \repr1, \repr1)^{0}_{0,0,0}+(\repr1, \repr1, \repr1)^{0}_{0,0,0}
\nonumber \\
&~& +(\repr2, \repr2, \repr1)^{2}_{2,0,0}+(\repr2, \repr2, \repr1)^{2}_{-2,0,0}
+(\repr2, \repr1, \repr1)^1_{1,5,0}+(\repr1, \repr2, \repr1)^{-1}_{-1,5,0}
\nonumber \\
&~& +(\repr2, \repr1, \repr1)^{-1}_{-1,-5,0}+(\repr1, \repr2, \repr1)^{1}_{1,-5,0}
+ (\repr1, \repr1, \repr1)^{0}_{0,0,0}
\nonumber \\
&~& +(\repr2, \repr1, \repr{\overline{4}})^{-1}_{1,1,9}+(\repr1, \repr2, \repr{\overline{4}})^{1}_{-1,1,9}
+(\repr2, \repr1, \repr4)^{1}_{-1,-1,-9}+(\repr1, \repr2, \repr4)^{-1}_{1,-1,-9}
\nonumber \\
&~& + (\repr1, \repr1, \repr{\overline{4}})^{2}_{0,-4,9}+(\repr1, \repr1, \repr4)^{2}_{0,4,-9}
+(\repr1, \repr1, \repr{15})^{0}_{0,0,0},
\label{Eq:decom80}
\eeqn
where representations/charges of $({\rm SU}(2)\sub{I}, {\rm SU}(2)\sub{II}, {\rm SU}(4))^{R_0}_{{\rm U}(1)_A, {\rm U}(1)_B, {\rm U}(1)_C}$
are shown at the indicated positions.
From \Eqref{Eq:decom80}, we find that the SM gauge bosons ($G_{\mu}$, $W_{\mu}$, $B_{\mu}$) and the Higgs doublets ($H_u$, $H_d$) are generated as
\beqn
A_M =\begin{cases}
 A_{\mu} \ni \begin{cases}
  \MyDecomC{1}{1}{8}{0}{0}{0}{0}{0} \Rightarrow G_{\mu} \\
  \MyDecomC{3}{1}{1}{0}{0}{0}{0}{0}+\MyDecomC{1}{3}{1}{0}{0}{0}{0}{0} \Rightarrow W_{\mu} \\
  \MyDecomC{1}{1}{1}{0}{0}{0}{0}{0}+\MyDecomC{1}{1}{1}{0}{0}{0}{0}{0},
  \MyDecomC{1}{1}{1}{0}{0}{0}{0}{0} \Rightarrow A^{(y)}_{\mu}, A^{(\rm IV)}_{\mu}
     \sim B_{\mu}, B^{\bot}_{\mu}
 \end{cases} \\
 A_{z} \ni \MyDecomC{2}{1}{1}{1}{5}{0}{0}{1} + \MyDecomC{1}{2}{1}{-1}{1}{9}{3}{1}
\Rightarrow H_u^* \\
 A_{\bar{z}} \ni \MyDecomC{1}{2}{1}{-1}{5}{0}{0}{-1} + \MyDecomC{2}{1}{1}{1}{1}{9}{3}{-1}
\Rightarrow H_d ,
\end{cases}
\label{Eq:bosons}
\eeqn
where representations/charges of 
$\Ls {\rm SU}(2)\sub{I},{\rm SU}(2)\sub{II}, {\rm SU}(3)\Rs_{{\rm U}(1)_A,{\rm U}(1)_B,{\rm U}(1)_C,{\rm U}(1)_D}^{R_0}$ are shown at the indicated positions, 
$A^{(y)}_{\mu}$ and $A^{({\rm IV})}_{\mu}$ are the gauge bosons of ${\rm U}(1)_{y}$ and ${\rm U}(1)_{\rm IV}$, respectively, and the gauge bosons $B_{\mu}$ and $B^{\bot}_{\mu}$ associated with ${\rm U}(1)_Y$ and ${\rm U}(1)_{Y^{\bot}}$ are given by linear combinations of them.
In \Eqref{Eq:bosons}, the assignment of two Higgs doublets is determined to be compatible with that of quarks originating from $\Psi_{+}(\repr{84})$, as in \Eqref{Eq:bosons-SU3}.

Using \Eqref{Eq:SU(9)GHU-RelsU(1)},
we can verify that these assignments of eqs.~\eqref{Eq:quarks+}, \eqref{Eq:quarks-}, \eqref{Eq:leptons+}, \eqref{Eq:leptons-} and \eqref{Eq:bosons}
match those of eqs.~\eqref{Eq:quarks+SU3}, \eqref{Eq:quarks-SU3}, \eqref{Eq:leptons+SU3}, \eqref{Eq:leptons-SU3} and \eqref{Eq:bosons-SU3}, respectively.

%%%%%%%%%%%%%%%%%%%%%%%%%%%%%%%%%%%%%%%%%%%%%%%%%%%%%%%%%%%%%%%%%%%%%%%%%%%%%

\end{document}